\documentclass[journal=mamobx,manuscript=article]{achemso}

\usepackage[version=3]{mhchem} 
\usepackage{xcolor}
\usepackage{chemfig}
\usepackage{braket}
\usepackage{amssymb,mathtools}
\usepackage{siunitx} 
\usepackage{subcaption}
\usepackage[final]{changes}

\usepackage{xr}
\makeatletter
\newcommand*{\addFileDependency}[1]{
  \typeout{(#1)}
  \@addtofilelist{#1}
  \IfFileExists{#1}{}{\typeout{No file #1.}}
}
\makeatother

\newcommand*{\myexternaldocument}[1]{
    \externaldocument{#1}
    \addFileDependency{#1.tex}
    \addFileDependency{#1.aux}
}

\myexternaldocument{si}

\newcommand\plainfootnote[1]{%
  \begingroup
  \renewcommand\thefootnote{}\footnote{#1}%
  \addtocounter{footnote}{-1}%
  \endgroup
}

\author{Robert H. Mei{\ss}ner}
\email{robert.meissner@tuhh.de}
\affiliation{Institute of Polymers and Composites, Hamburg University of Technology, Hamburg}
\alsoaffiliation{Institute of Materials Research, MagIC - Magnesium Innovation Centre, Helmholtz-Zentrum Geesthacht, Geesthacht}
\altaffiliation{These authors contributed equally to this work}

\author{Julian Konrad}
\affiliation{Department of Chemistry and Pharmacy, Computer Chemistry Center, Friedrich Alexander University Erlangen-Nürnberg, Erlangen}
\altaffiliation{These authors contributed equally to this work}

\author{Benjamin Boll}
\affiliation{Institute of Polymers and Composites, Hamburg University of Technology, Hamburg}

\author{Bodo Fiedler}
\affiliation{Institute of Polymers and Composites, Hamburg University of Technology, Hamburg}

\author{Dirk Zahn}
\affiliation{Department of Chemistry and Pharmacy, Computer Chemistry Center, Friedrich Alexander University Erlangen-N\"urnberg, Erlangen}

\title[Molecular simulation of thermosetting polymer hardening: reactive events enabled by controlled topology transfer]
  {Molecular simulation of thermosetting polymer hardening: reactive events enabled by controlled topology transfer}

\abbreviations{MD,QM/MM}

\begin{document}
\plainfootnote{This document is the unedited Author's version of a Submitted Work that was subsequently accepted for publication in Macromolecules, copyright \copyright American Chemical Society after peer review. To access the final edited and published work see http://dx.doi.org/10.1021/acs.macromol.0c02222.}

\begin{tocentry}




  \includegraphics[height=3.5cm]{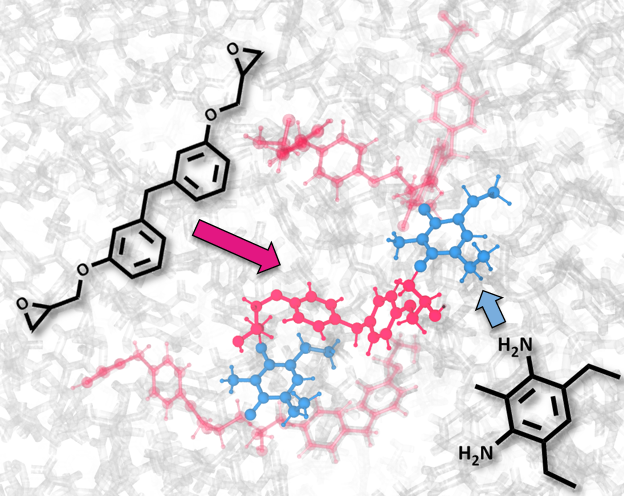}
\end{tocentry}

\begin{abstract}
  We present a 
  quantum mechanical / molecular mechanics (QM/MM) 
  to tackle chemical reactions with substantial molecular reorganization.  For this, molecular dynamics simulations with smoothly switched interaction models are used to suggest suitable product states, whilst a Monte Carlo algorithm is employed to assess the reaction likeliness subject to energetic feasibility. As a demonstrator, we study the cross-linking of bisphenol F diglycidyl ether (BFDGE) and 4,6-diethyl-2-methylbenzene-1,3-diamine (DETDA). The modeling of epoxy curing was supplemented by Differential Scanning Calorimetry (DSC) measurements, which confirms the degrees of cross-linking as a function of curing temperature. Likewise, the heat of formation and the mechanical properties of the resulting thermosetting polymer are found to be in good agreement with previous experiments.
\end{abstract}

\section{Introduction}

Thermosetting polymers may achieve high strength and stiffness, chemical resistance and offer simple processing. These advantages combined with low costs have enabled a wide manifold of industrial applications. 
For example, in structural elements used for harvesting wind energy, in automotive industries and for household devices, thermosetting polymers based on epoxy resins are used as matrix materials in fibre reinforced plastics of tailor-made mechanical properties. 

The yield strength\cite{Hobbiebrunken2007ExperimentalSpecimens,Sui2019ExtremeFibers} and the modes of plastic deformation\cite{Fiedler2001FailureLoading,Chevalier2018UnveilingThermosets} depend on loading conditions giving rise to plastic flow by cooperative chain conformation change and other mechanisms on the molecular level, including chemical bond reorganization.
Analogous to dislocation motion in metals, the experimental observation of deformations in amorphous plastics is practically impossible when it comes to the molecular level. 
The situation is even more precarious for the elucidation of formation mechanisms, that is the interplay of local reactive events and the overall bonding network constituting the bulk material.   
For this reason, molecular modelling and simulation is of crucial importance to provide profound understanding. 
To study molecular rearrangements occurring during thermoset hardening, plastic flow and fracture, Monte-Carlo and molecular dynamics (MD) simulations are particularly suited.  

Although pioneering studies \cite{Sundararaghavan2013MolecularTheory,Yang2014Coarse-grainedPolymer} have helped to understand elementary mechanisms of plasticity in thermosets and pre-polymer resins, a comprehensive understanding of the interplay of fundamental linking  processes and the collective organization of a polymer network is still missing. 
To enable the rationalization of materials response and macroscopic properties on the basis of molecular scale understanding, an obvious prerequisite is the realistic modelling of resin cross-linking.
As thermoset hardening is a non-equilibrium process, this calls for the explicit account of temperature and time-dependent behavior.     
Over the past decades several approaches have been published in order to model the cross-linkings of thermosetting polymers \cite{Gissinger2017ModelingSimulations,Varshney2008ArticleMaterial,Li2010MolecularPolymers,Vashisth2018AcceleratedPolymers,Jang2015ComparisonPolymers} and their interplay with mechanical properties\cite{Xin2015MolecularResin,Odegard2014PredictingReaxFF,Kallivokas2019MolecularDynamics,Li2015}.

For the accurate description of cross-linking reactions quantum mechanical (QM) calculations are required.
However, rigorous quantum approaches are unfeasible for systems comprising many thousands of atoms, as typically inevitable when modelling an amorphous thermosetting polymer.
To amend this limitation, quantum mechanical calculations may be coupled to classical molecular mechanics (MM), leading to QM/MM\cite{Warshel1976TheoreticalLysozyme} approaches that allow much larger time and length scales as a benefit of MM modelling the non-reactive part of the model, whilst reserving QM calculations to critical aspects requiring higher accuracy. 
However, despite the use of computationally efficient QM/MM or reactive MM approaches, the accessible time-scales of nano to micro seconds still prevent the direct molecular dynamics simulation of thermoset curing.  

To overcome such limitations, \citet{Jang2015ComparisonPolymers} proposed a Monte-Carlo based approach to step-wise cross-linking by energetically favorable reactions. 
This is particularly effective for modelling the starting period of thermoset curing, however gets increasingly ineffective at later stages.
Once larger degrees of cross-linking are established, the flexibility and the reorganization dynamics of the resin are drastically reduced and the ratio of exothermic versus endothermic reactive attempts in conventional Monte-Carlo approaches drops enormously. 
An interesting alternative to model accelerated cross-linking of epoxide systems has been recently introduced by \citet{Vashisth2018AcceleratedPolymers} using ReaxFF in combination with additional energetic kicks to help overcoming energetic barriers, and thus boost reactive events.
In principle this allows any degree of acceleration, however at the cost of describing molecular re-organization prior to cross-linking.  

The concept of the present work is to combine aspects of both of these approaches. 
Along this line, we control reactive attempts by means of Monte-Carlo selection criteria. 
However, to facilitate molecular reorganization needed for successful reactive attempts, we smoothly switch from one molecular topology to another by mixing the corresponding MM forces and energies.
On this basis, we aim at retaining the accuracy of QM/MM based Monte-Carlo evaluation of bond formation, whilst boosting network relaxation to enable also the efficient simulation of later stages of the curing process.

Ultimate benchmarks to this strategy are comparable degrees of cross-linking at different temperatures in both experiment and molecular simulation models. 
We hence relate our modelling approach to experimental characterization of cross-linking dynamics and mechanical properties. 
For the experiments, we use the commercially available thermosetting polymer of EPIKOTE{\texttrademark} Resin 862 from HEXION{\texttrademark} and the amino hardener ETHACURE\textsuperscript{\textregistered} 100 Plus Curative from ALBEMARLE\textsuperscript{\textregistered}. 
In turn, our simulation models focus on the main constituents of the experimental system, namely  bisphenol F diglycidyl ether (BFDGE) and 4,6-diethyl-2-methylbenzene-1,3-diamine (DETDA).

\section{Materials and Methods}

\subsection{Experimental characterization} \label{experimentalCharact}

\begin{figure*}[htb]
    \centering
    \includegraphics[width=\textwidth]{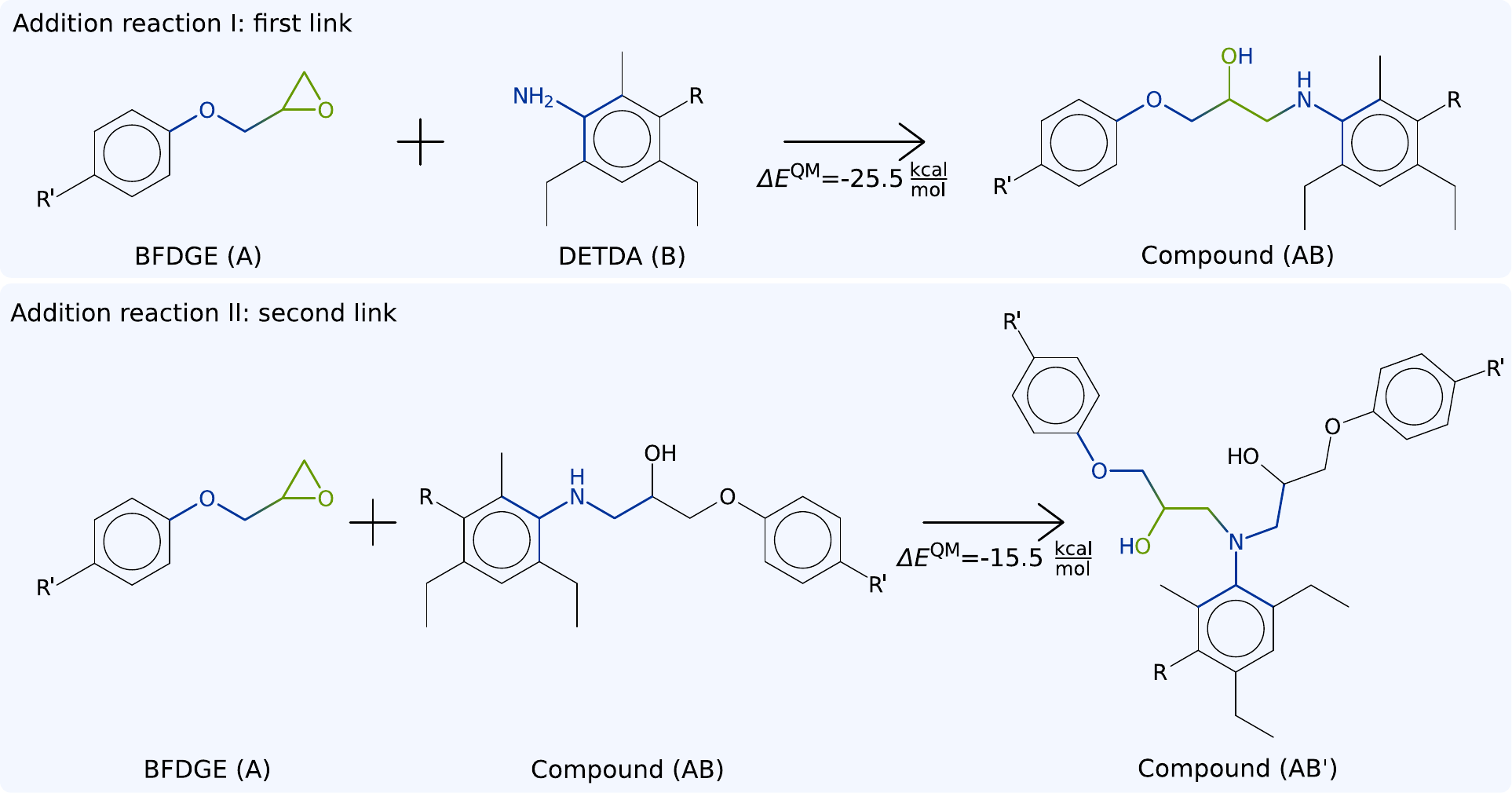}
    \caption{Chemical structures of BFDGE (upper left) and DETDA (upper middle) and illustration of the two step curing reaction of BFDGE and DETDA. Step I: single linked amine of DETDA to BFDGE (upper right). Step II: twofold linked amine of DETDA to two BFDGE molecules (lower right). Atoms that change their atom type are highlighted in green; atoms in which only the bond, angle, dihedral or improper type in which they are involved changes are highlighted in blue. Reaction energies are from QM calculations for R=NH$_2$ and R$^\prime$=H, respectively.}
    \label{fig:two_step_reaction}
\end{figure*}
\begin{figure}[htb]
    \centering
    \includegraphics[width=0.47\textwidth]{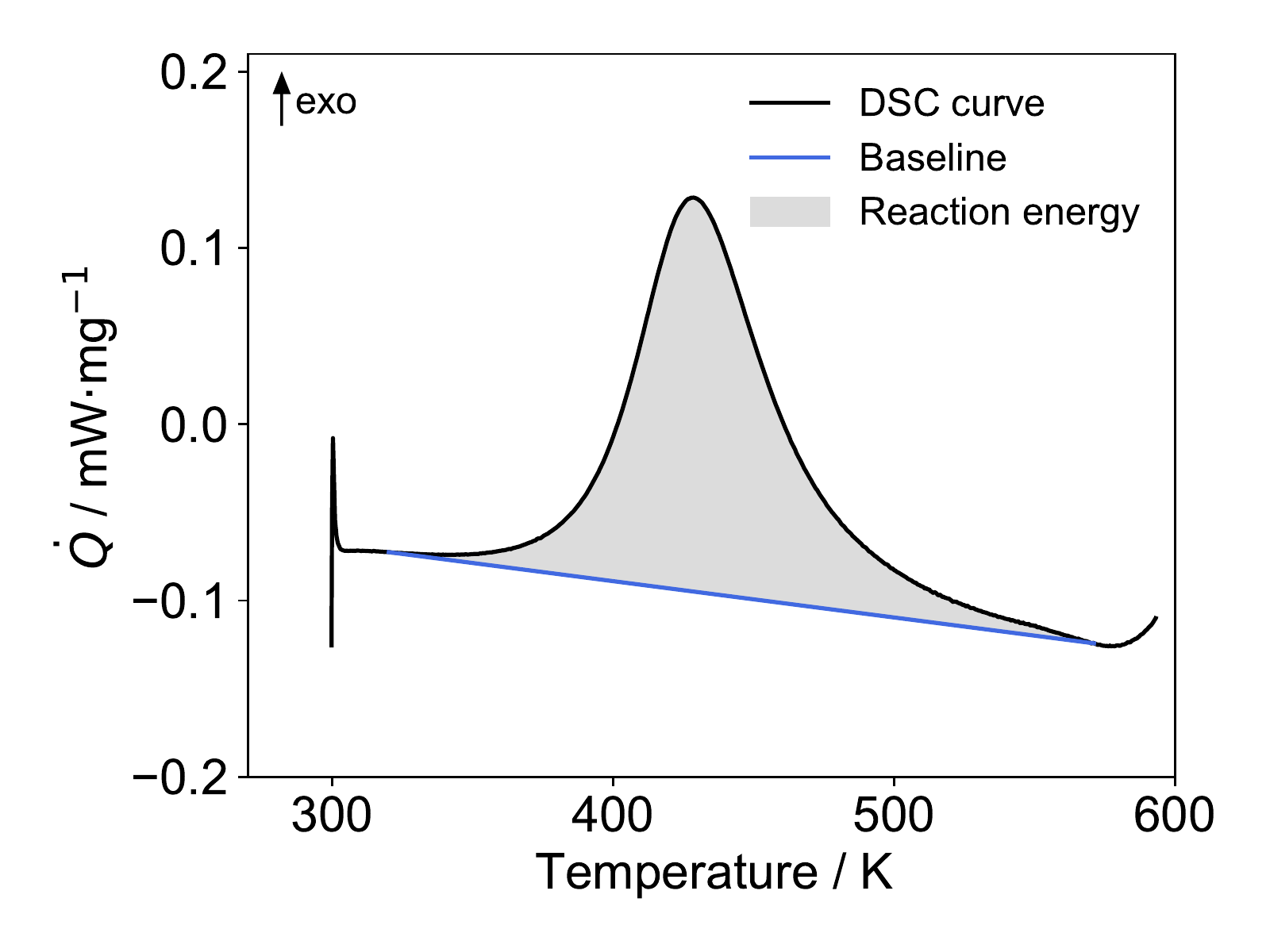}
    \includegraphics[width=0.47\textwidth]{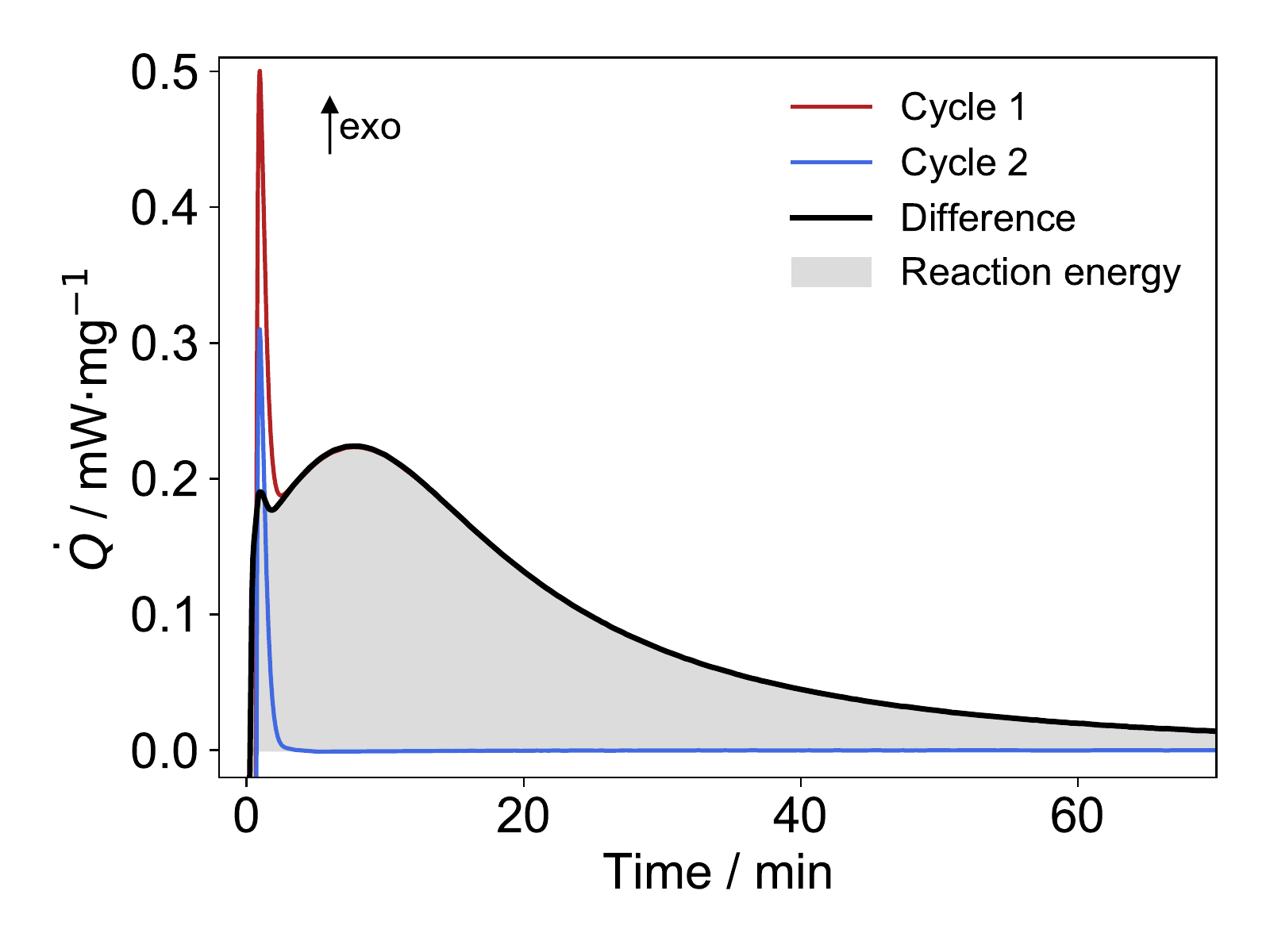}
    \caption{(left) DSC profile of the heat flux generated during thermoset hardening. The reaction energy (grey area) was calculated as the enclosed area between baseline and DSC signal. (right) Illustration of device specific overshoot. The resulting exothermic reaction energy is estimated as the integral of the corrected signal (grey area).}
    \label{fig:DSC_Fitting}
\end{figure}

%
%
As amine curing agent we use ETHACURE\textsuperscript{\textregistered} 100 Plus which is a composition of 3,5-diethyltoluene-2,4-diamine and 3,5-diethyltoluene-2,6-diamine at 75\,-\,81\,\% and 18\,-\,24\,\%, respectively. 
Moreover, it contains a small fraction of 0.5-3\,\% dialkylated \textit{m}-phenylenediamine. 
In what follows, this commercially available thermoset is simplified by its main constituents, BFDGE and DETDA as indicated in Figure~\ref{fig:two_step_reaction}. 
For the Differential Scanning Calorimetry (DSC) measurements a NETZSCH DSC 204 F1 Phoenix was used with pierced lit aluminum crucibles.
Before each measurement a freshly mixed two-component resin with a mass of 20.9$\pm$1.03\,mg was prepared ensuring equal experimental conditions.
The overall reaction energy of the thermoset hardening was averaged from measurements of five crucibles which were heated at a rate of 2\,K/min from room temperature at around 290\,K to 593\,K.
An exemplary DSC profile of the thermal ramp is shown in Figure~\ref{fig:DSC_Fitting} (left).
The peak between 360\,-\,580\,K stems from the cross-linking reactions, whereas the onset of thermoset decomposition is observed beyond 590\,K. 
The heat from the curing reaction was derived from the integral between the DSC curve and a linear baseline.

In the second test series, thermoset hardening is measured at different isothermal conditions adapting the measurement protocol proposed by Hernandez-Ortiz \textit{et al.}\cite{OrtizOsswald}.
The isothermal experiments were conducted at 380\,K, 420\,K and 460\,K. 
For each temperature, five samples were heated using a temperature ramp of 40\,K/min and cured isothermally until the thermoset curing was completed and the heat flow measured by the DSC converged to zero (cf. Fig.~\ref{fig:DSC_Fitting}).
Afterwards the sample is heated up to 590\,K to ensure a complete cure and cooled to room temperature again.
In a second identical cycle the so fully cured specimen was heated again to determine the device-specific temperature and heat flow overshoot of the DSC signal in the transition regime from heating to isothermal operation. 
The resulting second curve is subtracted from the first cycle to eliminate the artificial signal (cf. Fig.~\ref{fig:DSC_Fitting} (right)).
The area of the resulting curve was integrated in order to obtain isothermal exothermic reaction energies due to the cross-linking of the thermoset.

\subsection{QM description of epoxy bond formation}

The quantum chemical calculations were carried out with the Gaussian03 package at the B3LYP/6-311+G** level.
As starting points, geometry optimizations of the reactant species BFDGE (A) and DETDA (B), see also Figure~\ref{fig:two_step_reaction}, were performed. 
The first linking reaction, leading to compound AB (Fig.~\ref{fig:two_step_reaction}, upper panel), was induced by approaching A and B such that the epoxy group undergoes C-N bond formation and proton transfer from the amine at the same time. 
Compound AB was then subjected to geometry optimization and the reaction energy was found as -25.5 kcal/mol.

Likewise, the second linking reaction was characterized by addition of BFDGE (A) to compound AB, followed by structural relaxation, as illustrated in Figure~\ref{fig:two_step_reaction}, lower panel. 
The reactivity of the -NH- group in compound AB is substantially lower than that of species B and the reaction energy of the second linking of amine to epoxy amounts to only -15.5 kcal/mol. 
From the structural inspection of the resulting AB' compound we conclude that the reduction in bond formation energy may be rationalized from sterical hindering and hence deformation of the amine moiety in AB'.

\subsection{QM/MM based MC/MD simulations}

While the beforehand discussed QM characterization is limited to individual bond formation steps, the role of the entire polymer network shall be described by MM methods.
As the thermoset system in principle offers a large number of possible curing reaction pathways, we employ a Monte-Carlo (MC) strategy to probe reactions of nearby candidates. 
For this, we sample the MM energy of the polymer network before/after a possible polymerization step, and implement a QM/MM type approach to provide estimates of the reaction energy. 
Therein, the QM essentially accounts for epoxy-hardener reaction I (Fig.~\ref{fig:two_step_reaction}, upper panel) and II (Fig.~\ref{fig:two_step_reaction}, lower panel) whereas the network strain, embedding thermoset molecules etc. are all described by MM :
\begin{align}
\Delta E_\mathrm{I} &= 
    E_\mathrm{AB}^\mathrm{qm}-(E_\mathrm{A}^\mathrm{qm}+E_\mathrm{B}^\mathrm{qm}) 
    - E_\mathrm{AB}^\mathrm{mm}+(E_\mathrm{A}^\mathrm{mm}+E_\mathrm{B}^\mathrm{mm}) \nonumber\\
    & = (-25.5^\mathrm{qm} \nonumber + 38.8^\mathrm{mm}) \, \mathrm{kcal}\cdot\mathrm{mol}^{-1}  \, \nonumber
\end{align}
and
\begin{align}
\Delta E_\mathrm{II} &= 
    E_\mathrm{AB^\prime}^\mathrm{qm}-(E_\mathrm{AB}^\mathrm{qm}+E_\mathrm{A}^\mathrm{qm}) 
    - E_\mathrm{AB^\prime}^\mathrm{mm}+(E_\mathrm{AB}^\mathrm{mm}+E_\mathrm{A}^\mathrm{mm}) \nonumber \\
    & = (-15.5^\mathrm{qm} \nonumber + 42.1^\mathrm{mm}) \, \mathrm{kcal}\cdot\mathrm{mol}^{-1}\,.\nonumber
\end{align}
$E_\mathrm{A}^\mathrm{qm}$ and $E_\mathrm{B}^\mathrm{qm}$ are the potential energies of a monomeric resin BFDGE and hardener molecule DETDA, respectively, and $E_\mathrm{AB}^\mathrm{qm}$ and $E_\mathrm{AB^\prime}^\mathrm{qm}$ are the potential energies of the single or double bonded amino group to epoxides (cf. Fig.~\ref{fig:two_step_reaction}).
The $\Delta E_\mathrm{I}$ and $\Delta E_\mathrm{II}$ terms hence represent QM-based corrections to the MM description of forming isolated AB and AB' moieties. To account for the different embedding within the forming polymer network, the overall reaction energy is taken as:
\begin{equation}
  \Delta E =  \langle E_\mathrm{product}^\mathrm{mm} \rangle - \langle E_\mathrm{reactant}^\mathrm{mm} \rangle + 
  \begin{cases}
    \Delta E_\mathrm{I} \quad \text{or}\\
    \Delta E_\mathrm{II}
  \end{cases}
  \label{eqn:qm_correction}
\end{equation}
where $\langle E_\mathrm{reactant}^\mathrm{mm} \rangle$ and $\langle E_\mathrm{product}^\mathrm{mm}\rangle$ denote the energy of the overall simulation model before and after a reactive event, respectively, and the QM/MM correction term is selected depending on single or double bonded amino group binding to epoxide.
For both states, equilibrated systems are subjected to energy sampling from 2.5\,ps sketches to average out vibrational fluctuations. 
%
%
Acceptance or rejection of an reaction attempt is controlled by MC simulation using the Metropolis acceptance probability according to:
\begin{equation}
    p = \min \left(1,e^{-\frac{\Delta E}{k_\mathrm{B}T}} \right) ,
    \label{eqn:mc_prob}
\end{equation}
where $T$ and $k_\mathrm{B}$ are the temperature and the Boltzmann constant, respectively.

If a reaction attempt is not accepted, the interaction models are smoothly switched back to the previous topology.
Apart from such amendment of unfavorable linking within the current MC steps, we do not consider bond dissociation of the already formed polymer network. 
While surely important for the analysis of material deformation and fracture\cite{Yang2014Coarse-grainedPolymer}, for the present study of thermoset curing we ignore the possibility of bond dissociation steps.  

A molecular topology in empirical force fields refers in general to all bonded and non-bonded information of atomic species of a given molecule.
This encompasses all parameters in molecular mechanics necessary for the description of covalent bond lengths, valence angles and dihedrals as well as for the definition of molecular topology dependent atom types for van der Waals and Coulomb interactions.  
OPLS-AA topologies and partial charges on the semi-empirical 1.14*CM1A-LBCC level\cite{Dodda20171.14CM1A-LBCC:Simulations} for monomeric, single and double cross-linked bisphenol F diglycidyl ether (BFDGE) and 4,6-diethyl-2-methylbenzene-1,3-diamine (DETDA) are used to describe the MM part of the molecules using the online tool LigParGen\cite{Dodda2017LigParGenLigands}.
It generates from simple SMILES strings topologies, partial charges and an energetically minimized structure. 
However, resulting structures from LigParGen are not fully charge neutral (although up to 10\textsuperscript{-6}\,$e$) and some of the symmetric atoms do not possess the same charge.
Appropriate small compensating charges are thus added to/removed from the atom type with the highest absolute charge, where necessary, and charges are averaged for equivalent atoms to ensure charge neutrality of the molecules.
This is crucial to ensure consistent energy levels during the MD/MC simulation procedure. 

It should be noted that in the experiments the thermosetting polymer will undergo mixing and cross-linking at the same time.
When discussing the DSC experiments, we explicitly discriminate the heat of initial mixing from that of curing reactions. 
In turn, we start the molecular simulations from randomized starting configurations using moltemplate and PACKMOL\citep{Martinez2009PACKMOL:Simulations}.
Consequently, the heat of formation as sampled from simulation is set to zero for 0\% curing. 
The initial systems are unstable and may not fully relax, however before considering curing reactions we allowed 0.2\,ns runs which was found sufficient for the system to adapt local interactions and overall model density. This is followed by a short 0.05\,ns run at constant volume to sample the radial pair distribution functions of uncured thermoset mixtures.

The curing was modelled in an in the NpT ensemble at 1\,atm pressure at six different temperatures (260\,K, 300\,K, 340\,K, 380\,K, 420\,K and 460\,K) which have been as well used in the MC step.
During the smooth transfer of topology and the following equilibration, a relatively small timestep of 0.5\,fs was used in the MD simulations.

\subsection{Smooth transfer of topology}
\begin{figure*}
    \centering
    \includegraphics[width=0.47\textwidth]{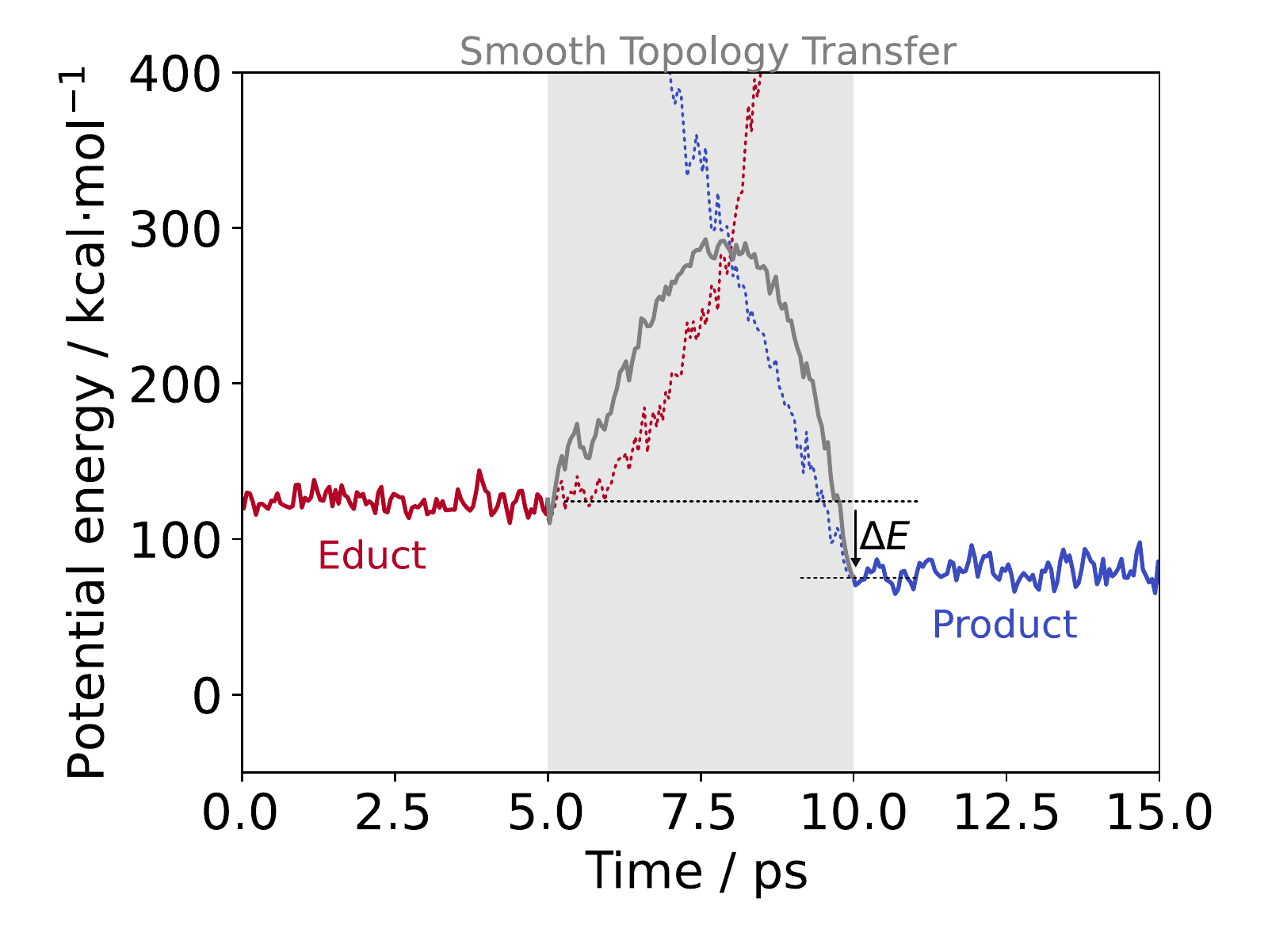}
    \includegraphics[width=0.47\textwidth]{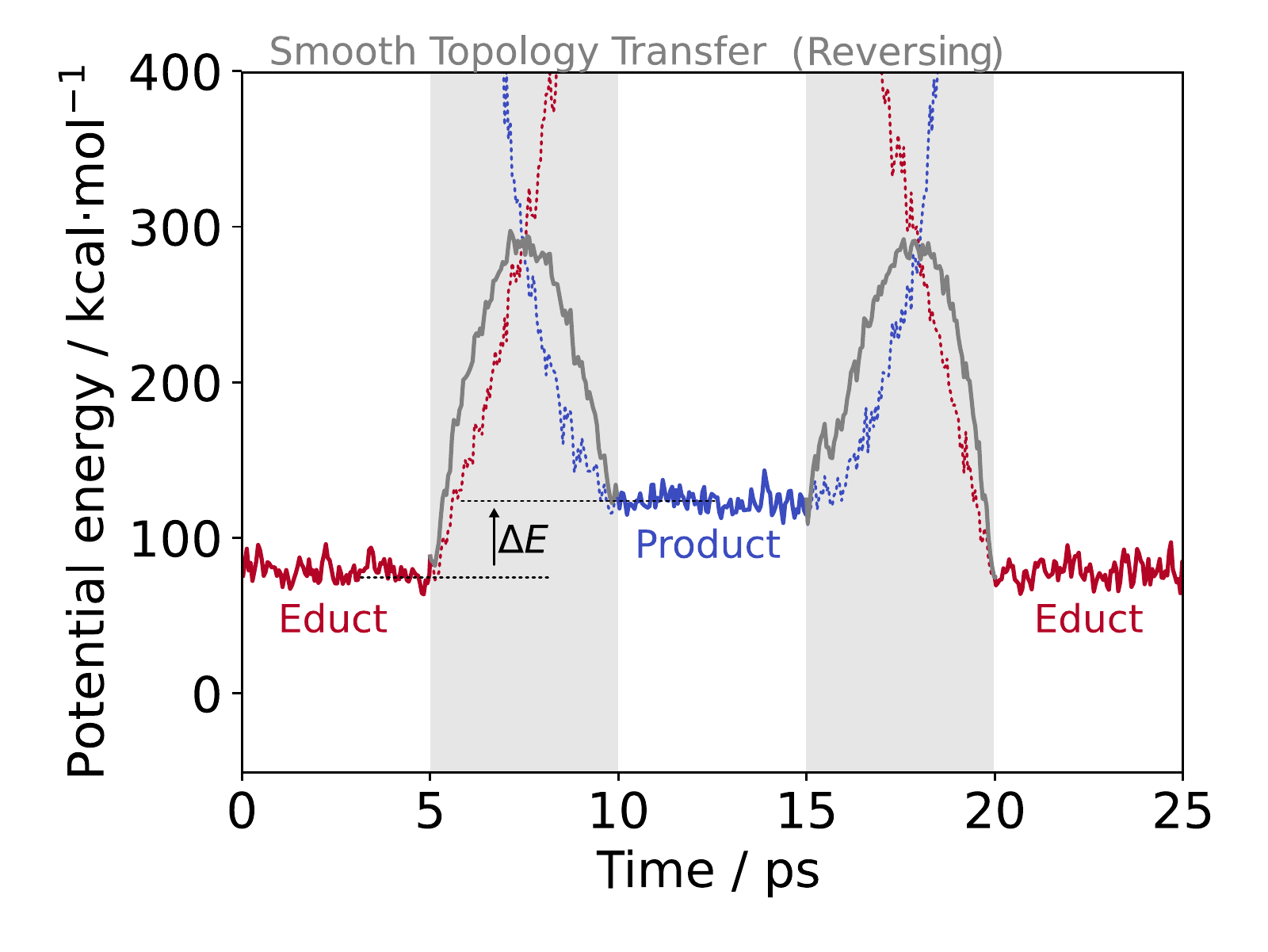}
    \caption{Potential energy as functions of time for (left) successful and (right) unsuccessful reaction attempts. The energy of the educt and product topologies are shown in red and blue, respectively, whilst grey indicates the transition in-between. In case the MC algorithm rejects a reaction attempt, we gradually reverse the binding topology of the system. }
    \label{fig:topo_example}
\end{figure*}

The linking reactions give rise to re-arrangement of both the local binding sites and the nearby polymer network.
For this reason, immediate switching of the interaction potential to the product state would often lead to unfavorable configurations--which may even involve molecular intersections that would provoke numerical instability. %
In order to efficiently suggest suitable candidate structures for linking reactions, we therefore perform MD simulations with smooth switching from one molecular topology to another, hence allowing network relaxation during gradual  forming of chemical bonds (cf. Fig.~\ref{fig:topo_example}).
When changing from one chemical topology to another,  the potential energy terms of several  bonds, angles, dihedrals and impropers have to be either dimmed off or ramped up. 
To provide continuous forces, this is implemented by a continuous switching function to smoothly transform the model topologies within a given time interval ($t_\mathrm{0}, t_\mathrm{end}$):
\begin{equation}
    s(t) = 1.0 - \frac{(t_\mathrm{end}^2 - t^2)^2 (t_\mathrm{end}^2+2t^2-3t_0^2)}{(t_\mathrm{end}^2 - t_0^2)^3} \, .
\end{equation}
Thus, for a linking reaction introduced for a configuration $\mathbf{X}(t)=(\mathbf{r}_1(t),\ldots,\mathbf{r}_N(t))$, the change in interaction energy is given by:
\begin{equation}
  E^\mathrm{topo}_\mathrm{reactant\rightarrow product}(\mathbf{X}(t),t) = s(t)\cdot \left[E_\mathrm{product}^\mathrm{mm}\left(\mathbf{X}(t)\right) - E_\mathrm{reactant}^\mathrm{mm}\left(\mathbf{X}(t)\right)\right] \, ,
  \label{eqn:smooth_topo_eng}
\end{equation}
whereas the corresponding forces $\vec{F}_j(t)$ acting on the atoms $j$ are:
\begin{align}
    \vec{F}_j(\mathbf{X}(t),t) = & -\nabla_j \big[E_\mathrm{reactant}^\mathrm{mm}(\mathbf{X}(t)) + E^\mathrm{topo}_\mathrm{reactant\rightarrow product}(\mathbf{X}(t),t)\big] \nonumber \\
    = & - s(t) \cdot \nabla_j E_\mathrm{product}^\mathrm{mm}(\mathbf{X}(t)) - \big(1-s(t)\big) \cdot \nabla_j E_\mathrm{reactant}^\mathrm{mm}(\mathbf{X}(t)) \nonumber \\
    = & s(t) \cdot F_j^\mathrm{product} + \big(1-s(t)\big) \cdot F_j^\mathrm{reactant}
    \label{eqn:smooth_topo_forces}
\end{align}
Where $F^\mathrm{reactant}_j$ and $F^\mathrm{product}_j$ are the forces derived from the MM models of the reactant and product state topology, respectively. 
After such transition, the system is described by the target topology only and the previous topology is fully removed from the simulation model. 
The form of the switching function affects only the dynamics of the topology transition and could be actually of any form--as long as it is a continuous function switching between zero and one.
With the present approach, we find that spikes in the energies or forces are avoided, hence allowing comparable fast topology transformations. 
It is worth mentioning that the smooth transfer of topology algorithm in almost all cases requires two force evaluations per time step. 
This downside is however overcompensated by much faster relaxation of the systems as compared to other approaches, e.g. a recently introduced method from \citet{Gissinger2017ModelingSimulations}, where instabilities in the simulation are compensated using a restriction of the maximum possible distance an atom can move in a single timestep. 
A custom implementation of the smooth transfer of topology method has been implemented in the Large-scale Atomic/Molecular Massively Parallel Simulator (LAMMPS)\citep{lammps} and is made available on GitHub or by request. 
Our implementation is moreover based on the extensive use of the \textit{python} interface supplied with LAMMPS, which enabled effectively the QM/MM 
simulations.

\section{Results and Discussion}

\subsection{Cross-linking Simulations}

%

\begin{figure}[htb]
    \centering
    \includegraphics[width=0.47\textwidth]{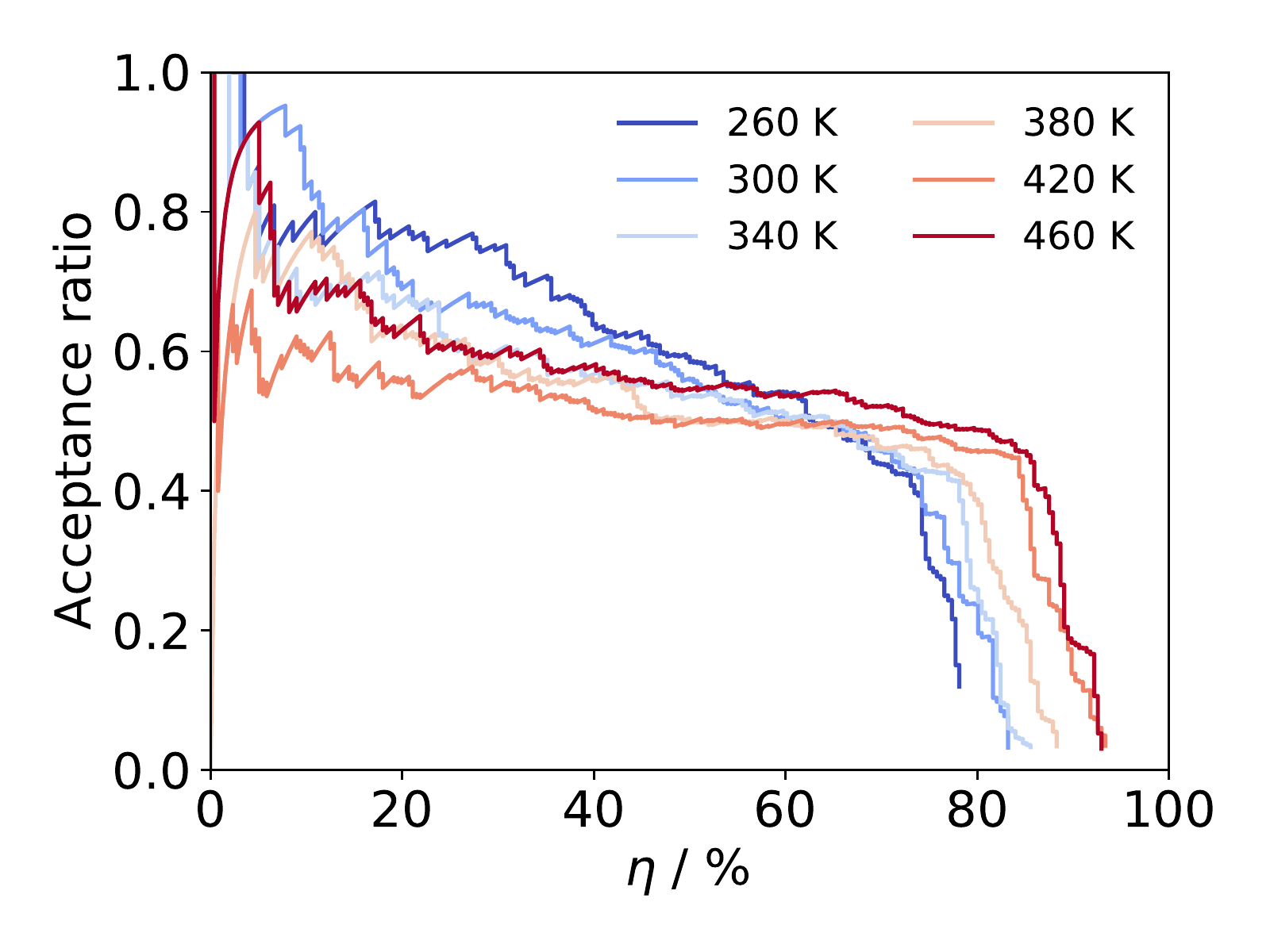}
    \caption{Monte-Carlo acceptance ratio with increasing degree of thermoset cure at different temperatures.}
    \label{fig:mc_convergence}
\end{figure}

\begin{figure}[htb]
    \includegraphics[width=1.0\textwidth]{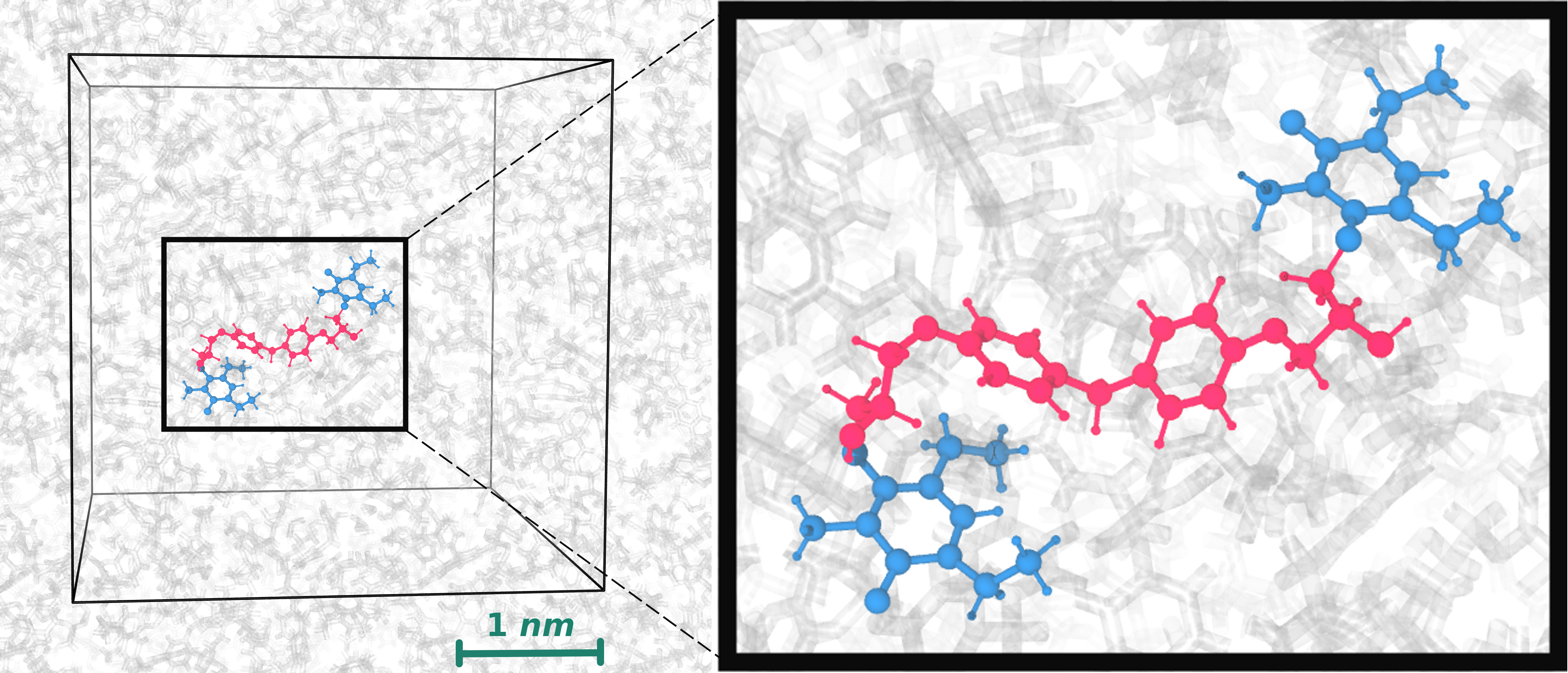}
    \caption{Illustration of the simulation cell of the thermoset at 93\% curing. The close-up illustrates two cross-links between BFDGE (red) and DEDTA (blue) moieties}
    \label{fig:cl_239_box}
\end{figure}

The combined Monte-Carlo/molecular dynamics simulations are initiated from fully mixed reactants which were allowed to relax at 1 atm pressure and the given temperature before considering any curing reaction. 
This initial model was also used to estimate a suitable distance cutoff for identifying eligible reaction partners. 
For this, the radial distribution functions (RDF) between epoxy and amine groups were evaluated at the six different temperatures considered for the thermoset curing (for details see SI).
Based on the RDFs, a 5\,\AA~cutoff was chosen as a reasonable common value to discriminate nearest-neighbor epoxy groups next to amine residues (cf. black dotted line in Fig.~\ref{fig:rdf_uncured}).
For the Monte-Carlo modelling of curing reactions, potential reaction partners are picked randomly from the (continuously updated) list of nearest-neighbors. Even for this pre-selection of candidates immediate switching from reactant to product state MM models would lead to a large number of failed attempts. However,  using our smooth topology transfer scheme we find that ps-scale molecular dynamics simulations can provide sufficient relaxation of the forming polymer network to suggest more realistic product states.
Attempts in which no reaction partner could be identified based on the distance delimiter are considered unsuccessful attempts in the Monte-Carlo simulation, hence anticipating strong energetic disfavoring without explicit evaluation. Instead, the system is propagated from MD simulations until a set of nearest-neighbors is found. 
In the course of the Monte-Carlo/MD simulation scheme, thermoset curing is thus modelled iteratively, as a combionation of attempting linking reactions and allowing network relaxation.
For the investigated range of temperatures between 260\,K and 460\,K, Figure~\ref{fig:mc_convergence} shows the evolution of the acceptance ratio ($N_\mathrm{accept}$/$N_\mathrm{attempts}$) of reaction attempts as functions of the degree of curing. 
On this basis, it appears that epoxy polymerization up to 60\% curing will depend on temperature only in terms of the reaction kinetics. The most obvious difference of the various curing temperature setups is seen at later stages of network formation, when the acceptance of reactive attempts sharply drop to zero. 
Depending on the temperature applied, the corresponding simulation models show final degrees of cross-linking ranging between 78\,\% and 93\,\%. However, inspecting the entire profile of the acceptance ratios, we argue that the different nature of polymer curing at different temperature is indicated from the decrease of successful linking attempts with increasing degree of curing which already starts at from the very beginning of network formation. 
Indeed, those binding partners (class I) which are already at nearest-neighbor distance when starting the Monte-Carlo iterations to model the curing process appear to react particularly exothermic. In turn, possible partners that require thermal fluctuations to occasionally fulfill the distance delimiter criterion (class II) tend to react less readily.
At low temperature, we therefore primarily find binding of class I partners - which explains the high acceptance rate at low degree of curing. However, once about 60\% curing is reached, further linking \textbf{must} involve class II type of partners. These, however, get increasingly jammed with increasing degree of curing and many of these eventually become inaccessible to low-temperature curing runs.  
On the other hand, at higher temperature there is a larger fraction of class II candidates in the pre-selection list throughout the entire curing run. As a consequence, network relaxation is considerably more intense in the high-temperature runs and much lesser amounts of possible reactants get trapped in the polymer network without a nearby partner.
An illustration of the model with the highest degree of curing is shown in Figure~\ref{fig:cl_239_box}, whereas bonding statistics of all runs are denoted in Table~\ref{tab:polymer_sim_results}. 
\begin{table*}[htb]
    \centering
    \begin{tabular}{ccccccccccccc}
        \hline
        $T_\mathrm{cure}$ & $\eta_\mathrm{sim}$ &         $\eta_\mathrm{exp}$ & \multicolumn{3}{c}{Amines (\%)} & \multicolumn{5}{c}{Functionality (\%)} & $Q_\mathrm{exp}$ & $Q_\mathrm{sim}$ \\
        (K) & (\%) & (\%) & Pri. & Sec. & Ter. & $f_0$ & $f_1$ & $f_2$ & $f_3$ & $f_4$ & (J/g) & (J/g) \\
        \hline
        260 & 78 & n/a & 8.6 & 27.3 & 64.1 & 1.7 & 6.8 & 8.5 & 35.6 & 47.5 & n/a & 344\\
        300 & 83 & n/a & 4.7 & 24.2 & 71.1 & 0.0 & 3.3 & 6.7 & 35.0 & 55.0 & n/a & 366\\
        340 & 86 & n/a & 2.3 & 24.2 & 73.4 & 0.0 & 0.0 & 6.6 & 37.7 & 55.7 & n/a & 376\\
        380 & 88 & 67 & 0. 8 & 21.9 & 77.3 & 0.0 & 1.6 & 10.9 & 20.3 & 67.2 & 295 $\pm$ 13 & 388\\
        420 & 93 & 87 & 0.0 & 13.3 & 86.7 & 0.0 & 0.0 & 4.7 & 17.2 & 78.1 & 383 $\pm$ \rlap{6}\phantom{13} & 410 \\
        460 & 93 & 95 & 0.8 & 12.5 & 86.7 & 0.0 & 1.6 & 3.1 & 17.2 & 78.1 & 418 $\pm$ \rlap{7}\phantom{13} & 409 \\
        \hline
    \end{tabular}
    \caption{Topological analysis of the polymer network models formed at different temperatures $T_\mathrm{cure}$. Primary, secondary and tertiary amines refer to NR\textsuperscript{1}H\textsubscript{2}, NR\textsuperscript{1}R\textsuperscript{2}H and NR\textsuperscript{1}R\textsuperscript{2}R\textsuperscript{3} moieties, respectively. To monitor the extend of unreacted groups we define $f_0$ as the fraction of DETDA with two primary amines. Each linking step is taken as an increase in the functionality index such that $f_4$ represents the fraction of DETDA with two tertiary amines.}
    \label{tab:polymer_sim_results}
\end{table*} %
The finally reached degree of curing of our simulation models studied at high temperature are found in quite nice agreement with the experiment. On the other hand, the incompleteness of linking reactions at lower temperature appears to be underestimated when using the smooth topology transfer simulation scheme.  
This is also reflected by comparing the reaction energies as computed from the QM/MM 
scheme with the DSC experiments as discussed in the following.

\subsection{Differential Scanning Calorimetry}

Two types of DSC experiments we performed. We first elucidated the heat of formation obtained for dynamically curing based on a temperature ramp as shown in Figure \ref{fig:DSC_Fitting}(left). For this setup, we found the largest heat of formation, namely 446\,$\pm$\,6.9\,Jg$^{-1}$, and in what follows we relate this to a curing degree of practically 100\%.
It is educative to compare the dynamically curing of the thermoset by a temperature ramp up to 600\,K, to DSC experiments performed at isothermal conditions--as illustrated for 380\,K on the right of Figure~\ref{fig:DSC_Fitting}. 
Indeed, the heat of polymer formation upon isothermally curing the thermoset at 380\,K, 420\,K and 460\,K are estimated to 295\,Jg$^{-1}$, 383\,Jg$^{-1}$ and 418\,Jg$^{-1}$, respectively.

\begin{figure}[htb]
    \centering
    \includegraphics[width=0.47\textwidth]{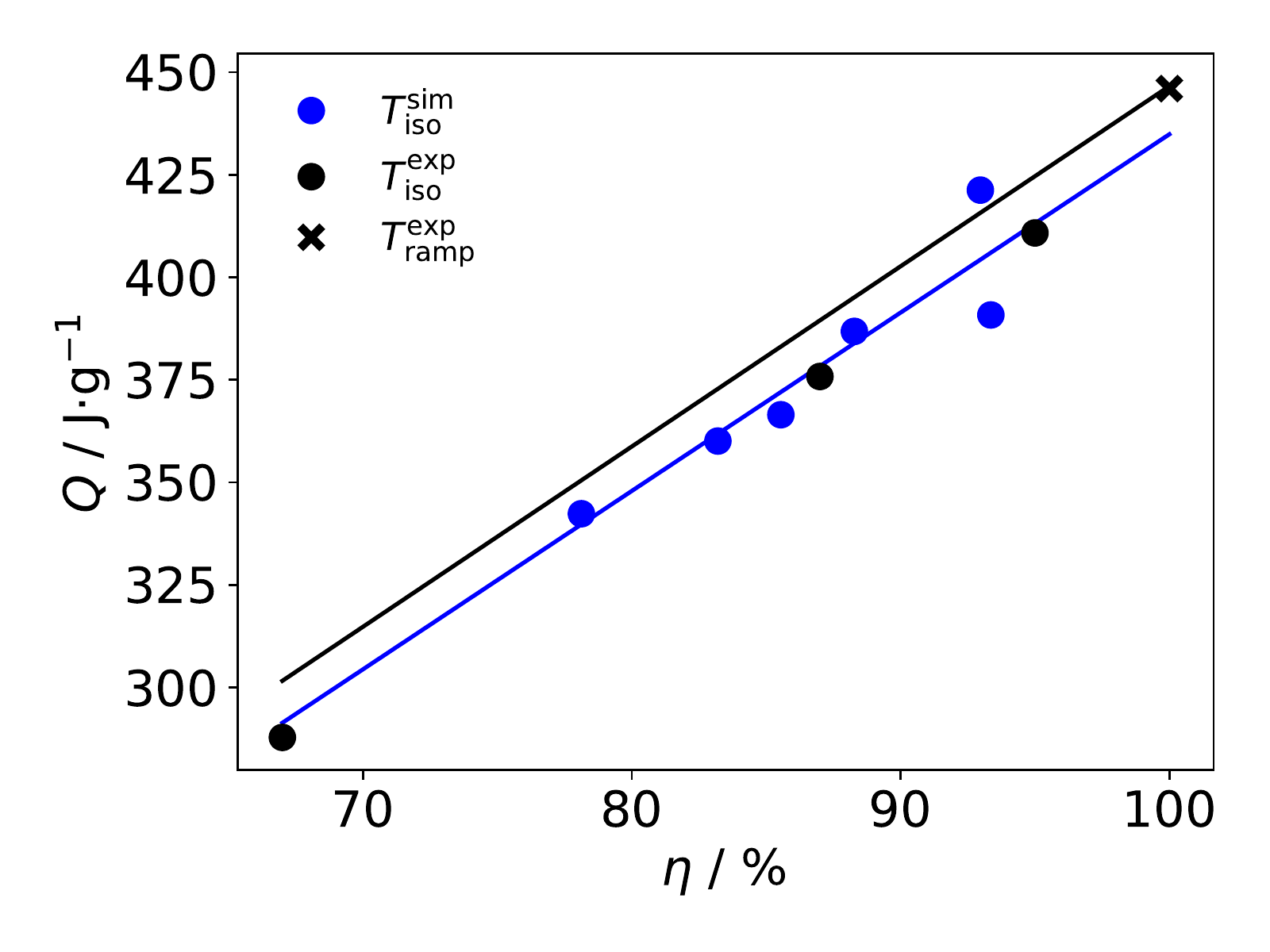}
    \caption{Heat of formation as obtained from DSC (black) and simulations (blue). The fully cured thermoset as identified from dynamical DSC measurements are indicated by the black cross. While the zero point energy of the simulations is defined such that that $Q(\eta=0)$ = 0\,J/g, the linear extrapolation of the DSC data leads to $Q(\eta=0)$ = (7$\pm$1\,J/g), hence indicating the error margins of the underlying integration procedure (cf. Fig.~\ref{fig:DSC_Fitting}) .}
    \label{fig:DSC_result}
\end{figure}

Due to the high latency of the commercial resin system, lower temperatures were not accessible in the isothermal measurements, as curing time scales would exceed weeks or even months.
The heat of formation as obtained from DSC experiments and our molecular simulations are shown as functions of finally reached degree of curing in Figure~\ref{fig:DSC_result}. For both, modelling and experiment, we find a linear dependence of the heat of formation with respect to the degree of curing. On this basis, we can derive the average reaction energy per extend of curing as 4.13\,$\pm$\,0.95\,Jg$^{-1}$/\% and 4.39\,$\pm$\,0.01\,Jg$^{-1}$/\%,
 thus reflecting a surprisingly close agreement between experiment and simulation, respectively.      
%
%
%

%
%

%
%

\subsection*{Elastic Properties} 

To characterize the bulk properties of our epoxide models with maximized cross-linking we performed constant-temperature, constant-pressure MD simulations of several 100\,ns. 
From this, occurence profiles of the fluctuations of the box volume $V$, simulation cell dimensions ($l_x$, $l_y$ and $l_z$), and angles ($\measuredangle (x,y)$, $\measuredangle (y,z)$ and $\measuredangle (z,x)$) were sampled and subjected to Boltzmann-statistics, hence providing elastic constants at 300\,K and 1\,atm from linear response theory.
The Young's moduli $Y$, shear moduli $G$ and the bulk modulus $K$ are thereby calculated by:
\begin{align}
  Y_x &=\frac{k_\mathrm{B}T}{(\Delta \sigma_{l_x})^2} \frac{\braket{l_x}^2}{\braket{V}} \nonumber\\
  G_{xy} &= \frac{k_\mathrm{B}T}{(\Delta \sigma_{\measuredangle (x,y)})^2\braket{V}} \nonumber \\
  K &= \frac{k_\mathrm{B}T}{(\Delta \sigma_\mathrm{V})^2}\braket{V} \nonumber
\end{align}
where the $\Delta \sigma$ are the width of Gaussian fits of the occurrence profiles of the underlying lengths, angles and the volume, respectively. 

\begin{figure}[htb]
  \begin{subfigure}[t]{0.3\textwidth}
    \includegraphics[width=1.0\textwidth]{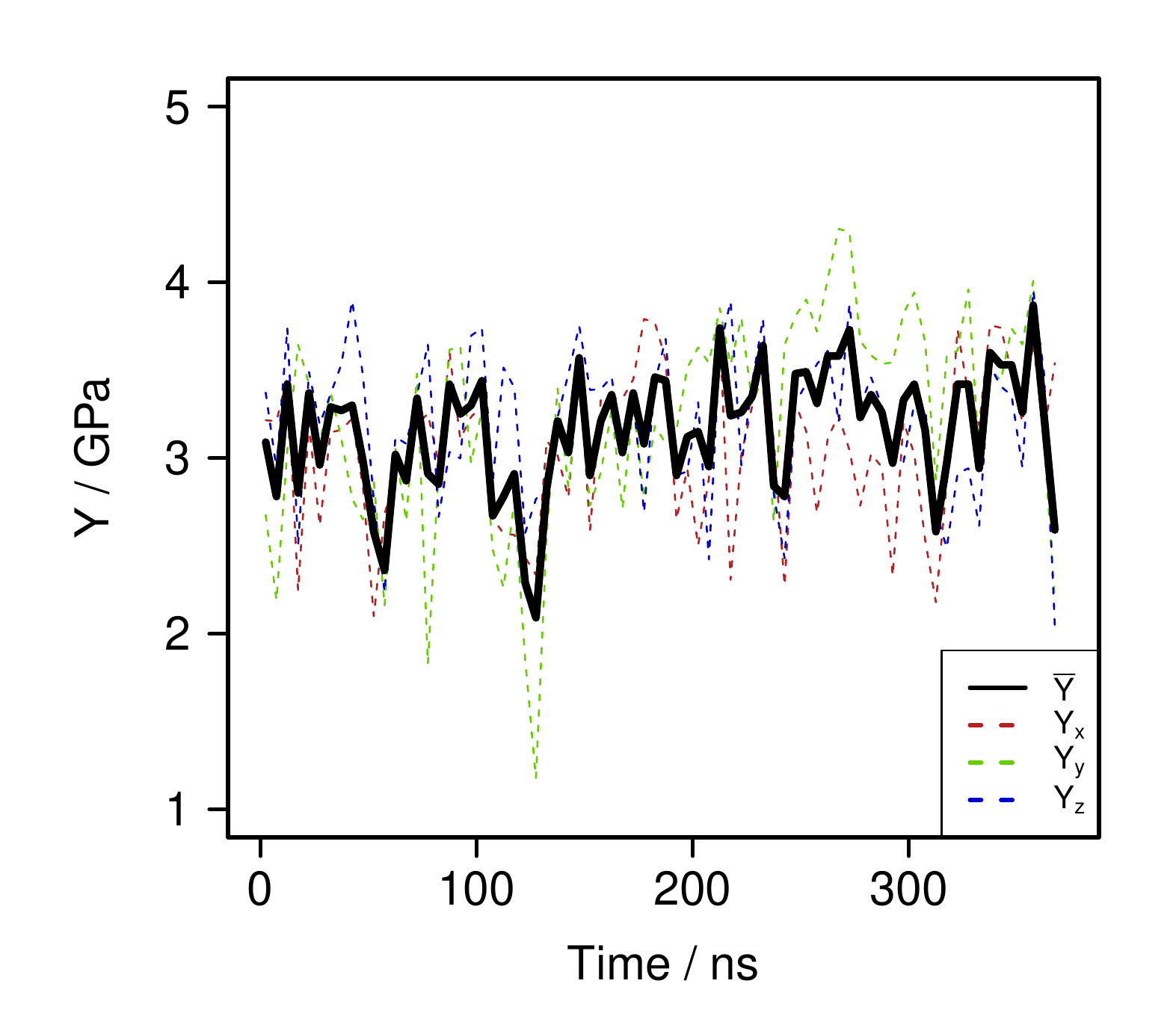}
    \subcaption{Young's}
    \label{fig:cl_239_young}
  \end{subfigure}
  \begin{subfigure}[t]{0.3\textwidth}
    \includegraphics[width=1.0\textwidth]{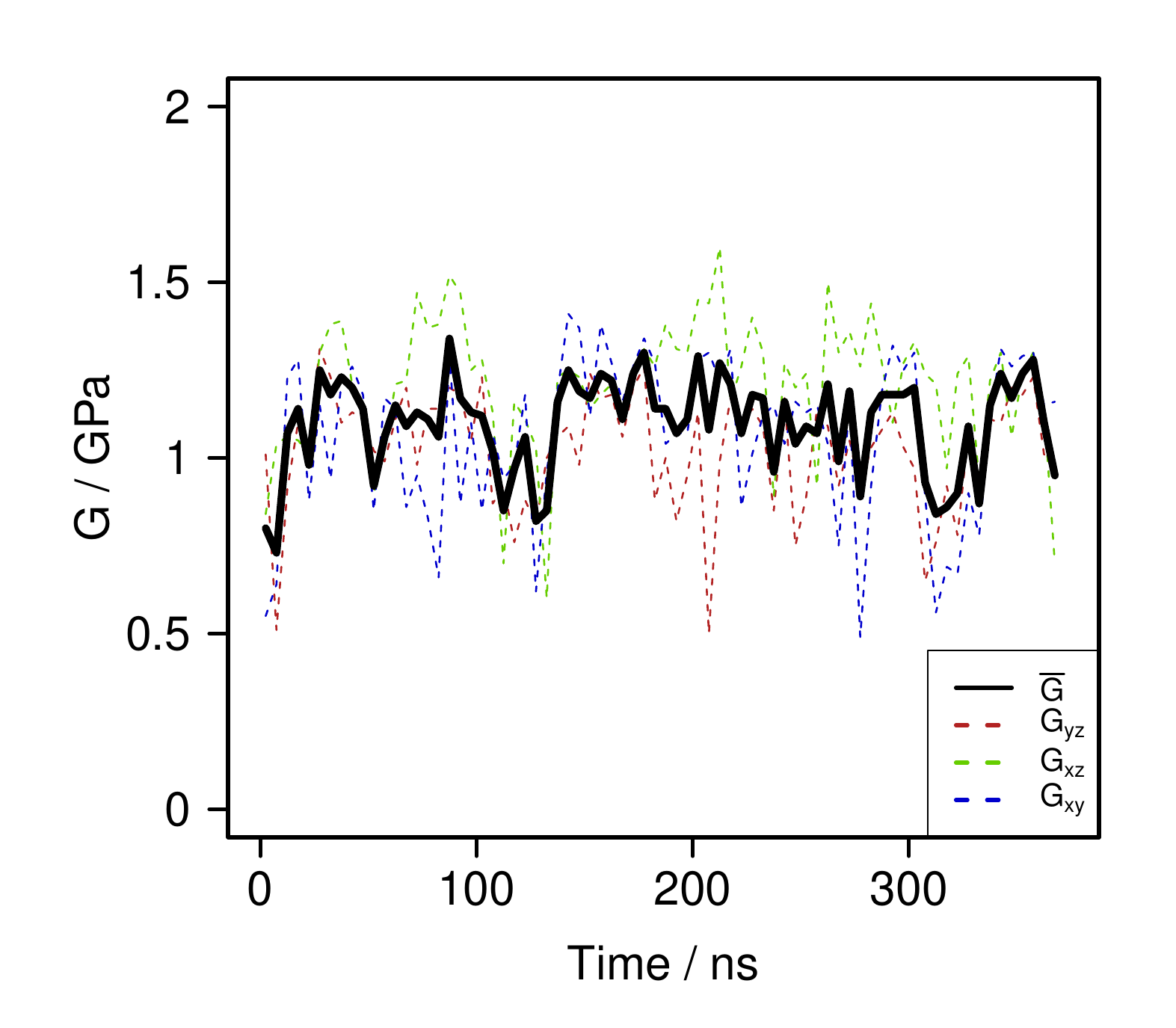}
    \subcaption{Shear}
  \end{subfigure}
  \begin{subfigure}[t]{0.3\textwidth}
    \includegraphics[width=1.0\textwidth]{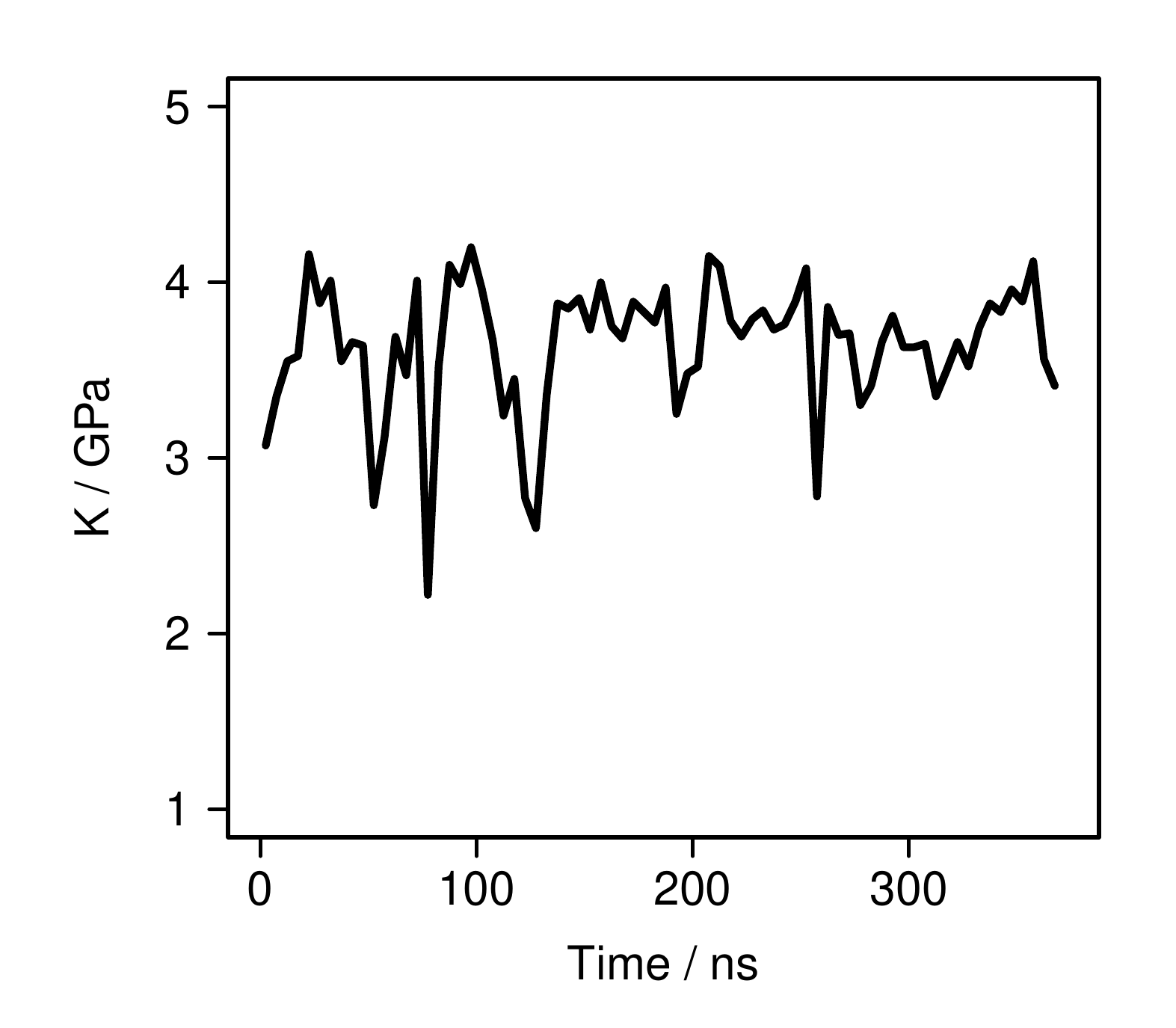}
    \subcaption{Bulk}
  \end{subfigure}
  \caption{Evolution of the elastic moduli over time and space directions. The moduli were calculated in 5\,ns intervals, overall time averages are provided in Table~\ref{tab_constant}. }
\end{figure}

While bulk epoxy resins are expected to be isotropic, we caution that our simulation cells are quite small compared to experimental setups and finite size effects need to be checked.
Indeed, we found that the Youngs and shear moduli taken along different directions show considerable differences.
Moreover, when performing time-dependent analyses based on series of 5\,ns subsamples, we observe fluctuations of the (local) elastic properties on the 100\,ns scale (Fig.~\ref{fig:cl_239_young}). 
Given the 10\,nm scale dimensions of our simulation model, such fluctuations stem from molecular twisting/distortion of individual links within the epoxy network without changing the connectivity. 
We note that this is a finite temperature effect that diminishes at 0\,K, hence demonstrating the importance of calculating the elastic properties of the epoxy models at the proper thermodynamic conditions.     

In Table~\ref{tab_constant} the time-averaged values for $Y$, $G$ and $K$ are provided, along with the Poisson number $\nu$--which was derived in different manners, namely using the elastic constants by $\nu(Y,G) = \frac{Y}{2G}-1$, $\nu(Y,K) = \frac{3K-Y}{6K}$ or $\nu(K,G) = \frac{3K-2G}{2(3K+G)}$. 

\begin{table}[htb]
  \centering
  \begin{tabular}{ l | l | l | l}
     & Our model & Kallivokas\citep{Kallivokas2019MolecularDynamics} model & Experiments \\
     & ($\eta$\,=\,93\%) & ($\eta$\,=\,92\%) & ($\eta$\,=\,100\%) \\
    \hline \hline
    $\langle Y \rangle$ / GPa& \textbf{3.16} $\pm$ 0.35 & 3.6 $\pm$ 0.3 & 1.66\,--\,2.96\textsuperscript{$\mathsection$}; 2.76\textsuperscript{$\ast$}; 2.46\textsuperscript{$\dagger$} \\
    $\langle G \rangle$ / GPa & \textbf{1.10} $\pm$ 0.14 & 1.3 $\pm$ 0.1 & 0.6\,--\,1.0\textsuperscript{$\mathsection$}; 0.95\textsuperscript{$\ddagger$}\\
    $\langle K \rangle$ / GPa & \textbf{3.64} $\pm$ 0.38 & 4.1 $\pm$ 0.2 & 2.9\textsuperscript{$\ddagger$} \\
    $\nu(\langle Y \rangle,\langle G \rangle)$ & 0.44 & & \\
    $\nu(\langle Y \rangle,\langle K \rangle)$ & 0.36 & & \\
    $\nu(\langle K \rangle,\langle G \rangle)$ & 0.36 & & \\
    $\langle\nu\rangle $ & \textbf{0.39} $\pm$ 0.05 & 0.40 $\pm$ 0.01 & 0.35\,--\,0.43\textsuperscript{$\mathsection$}
    \end{tabular}
    
   \textsuperscript{$\ddagger$}Kallivokas \textit{et al.}\citep{Kallivokas2019MolecularDynamics}; \textsuperscript{$\ast$} Sun \textit{et al.}\cite{Sun2008MechanicalComposites}; \textsuperscript{$\mathsection$}Littell \textit{et al.}\cite{Littell2008}; \textsuperscript{$\dagger$}Zhou \textit{et al.}
    
  \caption{Elastic constants as obtained from molecular simulations by smooth topology transformation (our model), straight bond enforcement as modelled by Kallivokas et al.\citep{Kallivokas2019MolecularDynamics} and experimental data taken from the literature.}
  \label{tab_constant}
\end{table}

Apart from comparing our models to the experiment, it is interesting to discuss the differences to other modelling data available. In the simulation study of Kallivokas et al. \citep{Kallivokas2019MolecularDynamics} the linking of binding partners chosen from a connectivity list was imposed in a simultaneous manner, hence allowing less relaxation to the polymer network. We argue that this should lead to stiffer networks as compared to our  models. Indeed, the elastic constants observed for the Kallivokas model are 10-20\% larger, despite the degree of curing being slightly smaller than in our model.

\subsection{Conclusions}

The complexity of cross-linking in epoxy resins calls for careful preparation of molecular simulation models. 
Clearly, \textit{ad hoc} asignment of the network structure can not be provided from intuition.
While clever algorithms for preparing connectivity lists offer the practically instantaneous implementation of a polymer network, subtle approaches should combine step-wise linking of molecules with relaxation of the overall system. 
However, mainly because of computational limitations, also our smooth topology transfer approach is not entirely free of bias. Indeed, a rigorous Monte-Carlo modelling must also consider network re-organization from bond dissociation and reformation--which would dramatically increase the length of our simulation runs.
%
%

%
The central benchmark criterion of simulation approaches to polymer formation is given by the degree of cross-linking achieved for the final simulation model. 
While commercial epoxy resins reach almost 100\,\% under optimal conditions,
our smooth topology transfer algorithm allowed for 93\,\% cross-linking as a top-score reached at 460\,K--which is in line with the corresponding isothermal DSC experiments.
Furthermore, our models reflect the decline of the curing degree when applying lower temperature, showing at least qualitative agreement with the corresponding experiments.
A particularly encouraging finding of our study is the nice agreement of the heat of formation observed as a function of the degree of curing. The assessment of reaction energies was enabled from a very simple and computationally efficient 
QM/MM modelling.
Indeed, only two types of reactions were discriminated, and both could be characterized from small subsystems. As the latter are isolated from the polymer network, only a single set of QM calculations is needed, whilst all effects from network distortion, long-range interactions etc. are described by inexpensive MM methods.
%

\begin{acknowledgement}

RHM and BB thanks the agency for science, research and gender equality of the city of Hamburg (BWFG, Beh\"orde f\"ur Wissenschaft, Forschung und Gleichstellung) for funding of the I$^3$-Lab VAM. Additional funding was provided by the Deutsche Forschungsgemeinschaft (DFG, German Research Foundation) Projektnummer 192346071 -- SFB 986 and Projektnummer 390794421 -- GRK 2462. JK and DZ acknowledge funding from the Graduiertenkolleg GRK 2423, 'FRASCAL', of the Friedrich-Alexander Universit\"at Erlangen-N\"urnberg.

\end{acknowledgement}

\begin{suppinfo}

Additional experimental DSC measurements and further structural analysis of the thermosetting polymer from the simulation is found in the Supporting Information (SI).

\end{suppinfo}

\bibliography{robeme,references}

\providecommand{\latin}[1]{#1}
\providecommand*\mcitethebibliography{\thebibliography}
\csname @ifundefined\endcsname{endmcitethebibliography}
  {\let\endmcitethebibliography\endthebibliography}{}
\begin{mcitethebibliography}{24}
\providecommand*\natexlab[1]{#1}
\providecommand*\mciteSetBstSublistMode[1]{}
\providecommand*\mciteSetBstMaxWidthForm[2]{}
\providecommand*\mciteBstWouldAddEndPuncttrue
  {\def\EndOfBibitem{\unskip.}}
\providecommand*\mciteBstWouldAddEndPunctfalse
  {\let\EndOfBibitem\relax}
\providecommand*\mciteSetBstMidEndSepPunct[3]{}
\providecommand*\mciteSetBstSublistLabelBeginEnd[3]{}
\providecommand*\EndOfBibitem{}
\mciteSetBstSublistMode{f}
\mciteSetBstMaxWidthForm{subitem}{(\alph{mcitesubitemcount})}
\mciteSetBstSublistLabelBeginEnd
  {\mcitemaxwidthsubitemform\space}
  {\relax}
  {\relax}

\bibitem[Hobbiebrunken \latin{et~al.}(2007)Hobbiebrunken, Fiedler, Hojo, and
  Tanaka]{Hobbiebrunken2007ExperimentalSpecimens}
Hobbiebrunken,~T.; Fiedler,~B.; Hojo,~M.; Tanaka,~M. \emph{Composites Part A:
  Applied Science and Manufacturing} \textbf{2007}, \emph{38}, 814--818\relax
\mciteBstWouldAddEndPuncttrue
\mciteSetBstMidEndSepPunct{\mcitedefaultmidpunct}
{\mcitedefaultendpunct}{\mcitedefaultseppunct}\relax
\EndOfBibitem
\bibitem[Sui \latin{et~al.}(2019)Sui, Tiwari, Greenfeld, Khalfin, Meeuw,
  Fiedler, and Wagner]{Sui2019ExtremeFibers}
Sui,~X.~M.; Tiwari,~M.; Greenfeld,~I.; Khalfin,~R.~L.; Meeuw,~H.; Fiedler,~B.;
  Wagner,~H.~D. \emph{Express Polymer Letters} \textbf{2019}, \emph{13},
  993--1003\relax
\mciteBstWouldAddEndPuncttrue
\mciteSetBstMidEndSepPunct{\mcitedefaultmidpunct}
{\mcitedefaultendpunct}{\mcitedefaultseppunct}\relax
\EndOfBibitem
\bibitem[Fiedler \latin{et~al.}(2001)Fiedler, Hojo, Ochiai, Schulte, and
  Ando]{Fiedler2001FailureLoading}
Fiedler,~B.; Hojo,~M.; Ochiai,~S.; Schulte,~K.; Ando,~M. \emph{Composites
  Science and Technology} \textbf{2001}, \emph{61}, 1615--1624\relax
\mciteBstWouldAddEndPuncttrue
\mciteSetBstMidEndSepPunct{\mcitedefaultmidpunct}
{\mcitedefaultendpunct}{\mcitedefaultseppunct}\relax
\EndOfBibitem
\bibitem[Chevalier \latin{et~al.}(2018)Chevalier, Brassart, Lani, Bailly,
  Pardoen, and Morelle]{Chevalier2018UnveilingThermosets}
Chevalier,~J.; Brassart,~L.; Lani,~F.; Bailly,~C.; Pardoen,~T.; Morelle,~X.~P.
  \emph{Journal of the Mechanics and Physics of Solids} \textbf{2018},
  \emph{121}, 432--446\relax
\mciteBstWouldAddEndPuncttrue
\mciteSetBstMidEndSepPunct{\mcitedefaultmidpunct}
{\mcitedefaultendpunct}{\mcitedefaultseppunct}\relax
\EndOfBibitem
\bibitem[Sundararaghavan and Kumar(2013)Sundararaghavan, and
  Kumar]{Sundararaghavan2013MolecularTheory}
Sundararaghavan,~V.; Kumar,~A. \emph{International Journal of Plasticity}
  \textbf{2013}, \emph{47}, 111--125\relax
\mciteBstWouldAddEndPuncttrue
\mciteSetBstMidEndSepPunct{\mcitedefaultmidpunct}
{\mcitedefaultendpunct}{\mcitedefaultseppunct}\relax
\EndOfBibitem
\bibitem[Yang and Qu(2014)Yang, and Qu]{Yang2014Coarse-grainedPolymer}
Yang,~S.; Qu,~J. \emph{Physical Review E - Statistical, Nonlinear, and Soft
  Matter Physics} \textbf{2014}, \emph{90}, 1--8\relax
\mciteBstWouldAddEndPuncttrue
\mciteSetBstMidEndSepPunct{\mcitedefaultmidpunct}
{\mcitedefaultendpunct}{\mcitedefaultseppunct}\relax
\EndOfBibitem
\bibitem[Gissinger \latin{et~al.}(2017)Gissinger, Jensen, and
  Wise]{Gissinger2017ModelingSimulations}
Gissinger,~J.~R.; Jensen,~B.~D.; Wise,~K.~E. \emph{Polymer} \textbf{2017},
  \emph{128}, 211--217\relax
\mciteBstWouldAddEndPuncttrue
\mciteSetBstMidEndSepPunct{\mcitedefaultmidpunct}
{\mcitedefaultendpunct}{\mcitedefaultseppunct}\relax
\EndOfBibitem
\bibitem[Varshney \latin{et~al.}(2008)Varshney, Patnaik, Roy, and
  Farmer]{Varshney2008ArticleMaterial}
Varshney,~V.; Patnaik,~S.~S.; Roy,~A.~K.; Farmer,~B.~L. \emph{Macromolecules}
  \textbf{2008}, \emph{41}, 6837--6842\relax
\mciteBstWouldAddEndPuncttrue
\mciteSetBstMidEndSepPunct{\mcitedefaultmidpunct}
{\mcitedefaultendpunct}{\mcitedefaultseppunct}\relax
\EndOfBibitem
\bibitem[Li and Strachan(2010)Li, and Strachan]{Li2010MolecularPolymers}
Li,~C.; Strachan,~A. \emph{Polymer} \textbf{2010}, \emph{51}, 6058--6070\relax
\mciteBstWouldAddEndPuncttrue
\mciteSetBstMidEndSepPunct{\mcitedefaultmidpunct}
{\mcitedefaultendpunct}{\mcitedefaultseppunct}\relax
\EndOfBibitem
\bibitem[Vashisth \latin{et~al.}(2018)Vashisth, Ashraf, Zhang, Bakis, and
  Duin]{Vashisth2018AcceleratedPolymers}
Vashisth,~A.; Ashraf,~C.; Zhang,~W.; Bakis,~C.~E.; Duin,~A. C. T.~V. \emph{The
  Journal of Physical Chemistry A} \textbf{2018}, \emph{122}, 6633--6642\relax
\mciteBstWouldAddEndPuncttrue
\mciteSetBstMidEndSepPunct{\mcitedefaultmidpunct}
{\mcitedefaultendpunct}{\mcitedefaultseppunct}\relax
\EndOfBibitem
\bibitem[Jang \latin{et~al.}(2015)Jang, Sirk, Andzelm, and
  Abrams]{Jang2015ComparisonPolymers}
Jang,~C.; Sirk,~T.~W.; Andzelm,~J.~W.; Abrams,~C.~F. \emph{Macromolecular
  Theory and Simulations} \textbf{2015}, \emph{24}, 260--270\relax
\mciteBstWouldAddEndPuncttrue
\mciteSetBstMidEndSepPunct{\mcitedefaultmidpunct}
{\mcitedefaultendpunct}{\mcitedefaultseppunct}\relax
\EndOfBibitem
\bibitem[Xin and Han(2015)Xin, and Han]{Xin2015MolecularResin}
Xin,~D.~R.; Han,~Q. \emph{Journal of Molecular Modeling} \textbf{2015},
  \emph{21}, 1--6\relax
\mciteBstWouldAddEndPuncttrue
\mciteSetBstMidEndSepPunct{\mcitedefaultmidpunct}
{\mcitedefaultendpunct}{\mcitedefaultseppunct}\relax
\EndOfBibitem
\bibitem[Odegard \latin{et~al.}(2014)Odegard, Jensen, Gowtham, Wu, He, and
  Zhang]{Odegard2014PredictingReaxFF}
Odegard,~G.~M.; Jensen,~B.~D.; Gowtham,~S.; Wu,~J.; He,~J.; Zhang,~Z.
  \emph{Chemical Physics Letters} \textbf{2014}, \emph{591}, 175--178\relax
\mciteBstWouldAddEndPuncttrue
\mciteSetBstMidEndSepPunct{\mcitedefaultmidpunct}
{\mcitedefaultendpunct}{\mcitedefaultseppunct}\relax
\EndOfBibitem
\bibitem[Kallivokas \latin{et~al.}(2019)Kallivokas, Sgouros, and
  Theodorou]{Kallivokas2019MolecularDynamics}
Kallivokas,~S.~V.; Sgouros,~A.~P.; Theodorou,~D.~N. \emph{Soft Matter}
  \textbf{2019}, \emph{15}, 721--733\relax
\mciteBstWouldAddEndPuncttrue
\mciteSetBstMidEndSepPunct{\mcitedefaultmidpunct}
{\mcitedefaultendpunct}{\mcitedefaultseppunct}\relax
\EndOfBibitem
\bibitem[Li and Strachan(2015)Li, and Strachan]{Li2015}
Li,~C.; Strachan,~A. \emph{Journal of Polymer Science Part B: Polymer Physics}
  \textbf{2015}, \emph{53}, 103--122\relax
\mciteBstWouldAddEndPuncttrue
\mciteSetBstMidEndSepPunct{\mcitedefaultmidpunct}
{\mcitedefaultendpunct}{\mcitedefaultseppunct}\relax
\EndOfBibitem
\bibitem[Warshel and Levitt(1976)Warshel, and
  Levitt]{Warshel1976TheoreticalLysozyme}
Warshel,~A.; Levitt,~M. \emph{Journal of Molecular Biology} \textbf{1976},
  \emph{103}, 227--249\relax
\mciteBstWouldAddEndPuncttrue
\mciteSetBstMidEndSepPunct{\mcitedefaultmidpunct}
{\mcitedefaultendpunct}{\mcitedefaultseppunct}\relax
\EndOfBibitem
\bibitem[Hern{\'a}ndez-Ortiz \latin{et~al.}(2012)Hern{\'a}ndez-Ortiz, Osswald,
  and Restrepo-Zapata]{OrtizOsswald}
Hern{\'a}ndez-Ortiz,~J.; Osswald,~T.; Restrepo-Zapata,~N. Modeling and analysis
  of cure kinetics of aliphatic epoxy resin with and without diffusion. ANTEC
  2012: 70th Annual Technical Conference of the Society of Plastics Engineers
  2012, Orlando, Florida, USA, 2-4 April 2012. 2012\relax
\mciteBstWouldAddEndPuncttrue
\mciteSetBstMidEndSepPunct{\mcitedefaultmidpunct}
{\mcitedefaultendpunct}{\mcitedefaultseppunct}\relax
\EndOfBibitem
\bibitem[Dodda \latin{et~al.}(2017)Dodda, Vilseck, Tirado-Rives, and
  Jorgensen]{Dodda20171.14CM1A-LBCC:Simulations}
Dodda,~L.~S.; Vilseck,~J.~Z.; Tirado-Rives,~J.; Jorgensen,~W.~L. \emph{Journal
  of Physical Chemistry B} \textbf{2017}, \emph{121}, 3864--3870\relax
\mciteBstWouldAddEndPuncttrue
\mciteSetBstMidEndSepPunct{\mcitedefaultmidpunct}
{\mcitedefaultendpunct}{\mcitedefaultseppunct}\relax
\EndOfBibitem
\bibitem[Dodda \latin{et~al.}(2017)Dodda, De~Vaca, Tirado-Rives, and
  Jorgensen]{Dodda2017LigParGenLigands}
Dodda,~L.~S.; De~Vaca,~I.~C.; Tirado-Rives,~J.; Jorgensen,~W.~L. \emph{Nucleic
  Acids Research} \textbf{2017}, \emph{45}, W331--W336\relax
\mciteBstWouldAddEndPuncttrue
\mciteSetBstMidEndSepPunct{\mcitedefaultmidpunct}
{\mcitedefaultendpunct}{\mcitedefaultseppunct}\relax
\EndOfBibitem
\bibitem[Mart{\'{i}}nez \latin{et~al.}(2009)Mart{\'{i}}nez, Andrade, Birgin,
  and Mart{\'{i}}nez]{Martinez2009PACKMOL:Simulations}
Mart{\'{i}}nez,~L.; Andrade,~R.; Birgin,~E.~G.; Mart{\'{i}}nez,~J.~M.
  \emph{Journal of Computational Chemistry} \textbf{2009}, \emph{30},
  2157--2164\relax
\mciteBstWouldAddEndPuncttrue
\mciteSetBstMidEndSepPunct{\mcitedefaultmidpunct}
{\mcitedefaultendpunct}{\mcitedefaultseppunct}\relax
\EndOfBibitem
\bibitem[Plimpton(1995)]{lammps}
Plimpton,~S. \emph{Journal of computational physics} \textbf{1995}, \emph{117},
  1--19\relax
\mciteBstWouldAddEndPuncttrue
\mciteSetBstMidEndSepPunct{\mcitedefaultmidpunct}
{\mcitedefaultendpunct}{\mcitedefaultseppunct}\relax
\EndOfBibitem
\bibitem[Sun \latin{et~al.}(2008)Sun, Warren, O’Reilly, Everett, Lee, Davis,
  Lagoudas, and Sue]{Sun2008MechanicalComposites}
Sun,~L.; Warren,~G.; O’Reilly,~J.; Everett,~W.; Lee,~S.; Davis,~D.;
  Lagoudas,~D.; Sue,~H.-J. \emph{Carbon} \textbf{2008}, \emph{46},
  320--328\relax
\mciteBstWouldAddEndPuncttrue
\mciteSetBstMidEndSepPunct{\mcitedefaultmidpunct}
{\mcitedefaultendpunct}{\mcitedefaultseppunct}\relax
\EndOfBibitem
\bibitem[Littell \latin{et~al.}(2008)Littell, Ruggeri, Goldberg, Roberts,
  Arnold, and Binienda]{Littell2008}
Littell,~J.~D.; Ruggeri,~C.~R.; Goldberg,~R.~K.; Roberts,~G.~D.; Arnold,~W.~A.;
  Binienda,~W.~K. \emph{Journal of Aerospace Engineering} \textbf{2008},
  \emph{21}, 162--173\relax
\mciteBstWouldAddEndPuncttrue
\mciteSetBstMidEndSepPunct{\mcitedefaultmidpunct}
{\mcitedefaultendpunct}{\mcitedefaultseppunct}\relax
\EndOfBibitem
\end{mcitethebibliography}


\providecommand{\latin}[1]{#1}
\providecommand*\mcitethebibliography{\thebibliography}
\csname @ifundefined\endcsname{endmcitethebibliography}
  {\let\endmcitethebibliography\endthebibliography}{}
\begin{mcitethebibliography}{2}
\providecommand*\natexlab[1]{#1}
\providecommand*\mciteSetBstSublistMode[1]{}
\providecommand*\mciteSetBstMaxWidthForm[2]{}
\providecommand*\mciteBstWouldAddEndPuncttrue
  {\def\EndOfBibitem{\unskip.}}
\providecommand*\mciteBstWouldAddEndPunctfalse
  {\let\EndOfBibitem\relax}
\providecommand*\mciteSetBstMidEndSepPunct[3]{}
\providecommand*\mciteSetBstSublistLabelBeginEnd[3]{}
\providecommand*\EndOfBibitem{}
\mciteSetBstSublistMode{f}
\mciteSetBstMaxWidthForm{subitem}{(\alph{mcitesubitemcount})}
\mciteSetBstSublistLabelBeginEnd
  {\mcitemaxwidthsubitemform\space}
  {\relax}
  {\relax}

\bibitem[Mart{\'{i}}nez \latin{et~al.}(2009)Mart{\'{i}}nez, Andrade, Birgin,
  and Mart{\'{i}}nez]{Martinez2009PACKMOL:Simulations}
Mart{\'{i}}nez,~L.; Andrade,~R.; Birgin,~E.~G.; Mart{\'{i}}nez,~J.~M.
  \emph{Journal of Computational Chemistry} \textbf{2009}, \emph{30},
  2157--2164\relax
\mciteBstWouldAddEndPuncttrue
\mciteSetBstMidEndSepPunct{\mcitedefaultmidpunct}
{\mcitedefaultendpunct}{\mcitedefaultseppunct}\relax
\EndOfBibitem
\end{mcitethebibliography}

\end{document}


\section{High temperature isothermal DSC measurements}

\begin{figure}
    \centering
    \includegraphics[width=0.6\textwidth]{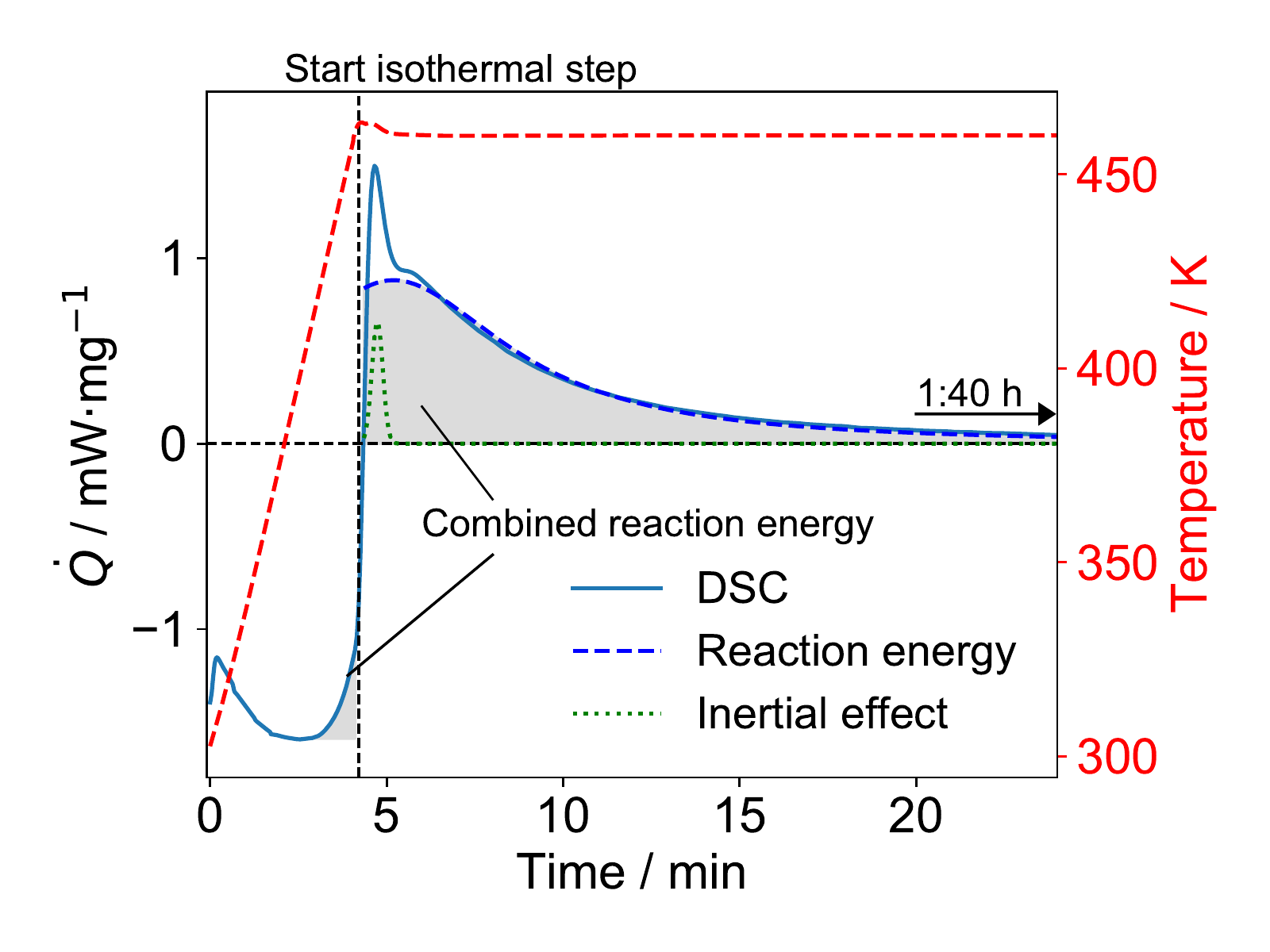}
    \caption{Estimation of exothermal heat generated during heating for curing isothermally a thermosetting polymer at 460\,K. The gray area under the green DSC curve left of the black dotted line (which denotes the heating phase) is assumed to be associated with the reaction heat generated during heating. The overshooting of the DSC heat flux (and temperature) after heating is completed is additionally shown and denoted as ``inertial effect''. See main text how the reaction heat is compensated for this.}
    \label{fig:iso_DSC_high_T_correction}
\end{figure}

In case of isothermal curing at 460\,K the temperature is far above the point where the reaction actually starts (360\,K\,--\,370\,K, see main text).
Thus, part of the polymerisation took place already during the heating phase.
The approach to estimate the additional contribution to the exothermal heat due to the heating is explained in Fig.~\ref{fig:iso_DSC_high_T_correction}.
It should be noted that this effect is negligible for isothermal curing below 420\,K ($<$1\,J/g), but becomes relevant for isothermal curing of the thermoset polymer at 460\,K ($\approx$10\,J/g). 

\section{Radial distribution function of an uncured thermoset}

\begin{figure}[htbp]
    \centering
    \includegraphics[width=0.6\textwidth]{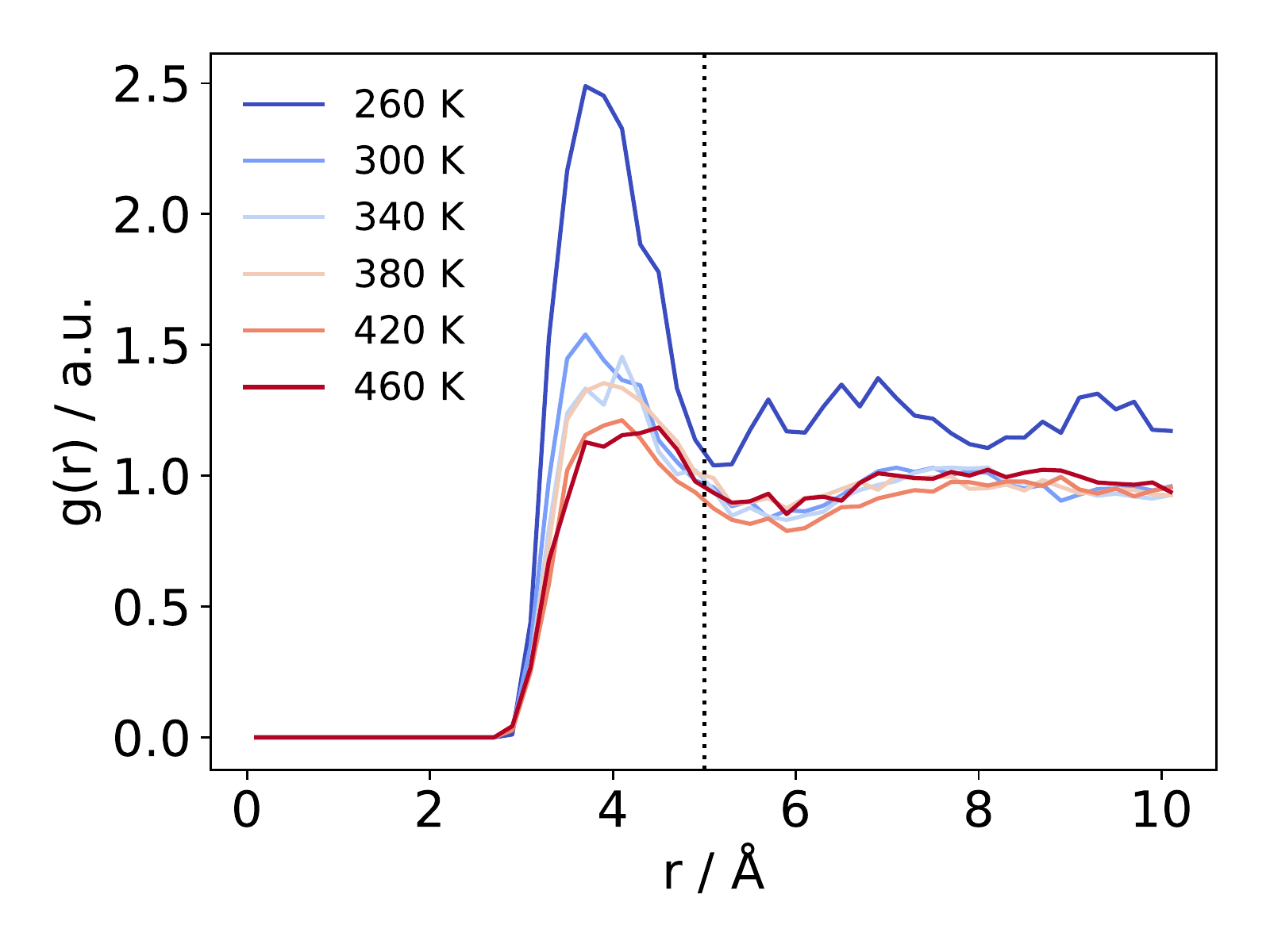}
    \caption{RDF $g(r)$ of the epoxy carbon edge in the resin and nitrogen in the amine group of the hardener. The dashed black line at $r_\mathrm{c}$=5\,\AA~marks the cutoff for finding reactive pairs.}
    \label{fig:rdf_uncured}
\end{figure}

The cutoff to determine eligible reaction partner is based on the Radial Distribution Function (RDF) for uncured thermosets at temperatures ranging between 260\,K\,-\,460\,K as shown in Fig.~\ref{fig:rdf_uncured}.
Before calculating the RDFs systems are equilibrated for 200\,ps in an NpT ensemble at their respective temperature and 1\,atm after randomly placing the molecules with packmol\cite{Martinez2009PACKMOL:Simulations} into a periodic box ensuring that the density is approximately correct.
RDFs shown in Fig.~\ref{fig:rdf_uncured} are obtained by averaging them in a subsequent NVT run over 50\,ps.
With increasing temperature a negligible shift of the first coordination peak to larger distances and a decrease in overall density at elevated temperatures (from 1.05\,g/cm$^{3}$ at 300\,K to 0.95\,g/cm$^{3}$ at 460\,K) was observed.
The system at 260\,K was in a rather amorphous but frozen state with a density of 0.84\,g/cm$^{3}$.
In addition, a decrease in peak area is observed with increasing temperature, indicating in general a lower amount of reactive epoxy groups around an amine or a higher mobility of both groups.
\bibliography{robeme}